\documentclass[11pt]{article}
\usepackage{graphicx}% Include figure files
\usepackage{dcolumn}% Align table columns on decimal point
\usepackage{bm}% bold math

\usepackage{fancyhdr}
\usepackage{amsmath} 
\usepackage{graphicx}
\usepackage{epsfig}
\usepackage{tabularx}
\usepackage{amsthm}
\usepackage{amsfonts}
\usepackage{amssymb}
\usepackage{endnotes}
\usepackage{colortbl}
\usepackage{makeidx}
\usepackage{bbold}
\usepackage{bbm}
\usepackage[samesize]{cancel}
\usepackage{mathtools}
\usepackage{color}
\usepackage{enumerate}
\usepackage{textcomp}
\usepackage{relsize}
\usepackage{enumitem} 
\usepackage{bbold}
\usepackage[top=2cm, bottom=2cm, left=2.cm , right=2.cm]{geometry}
\newcommand{\E}{\mathrm{E}}

\newtheoremstyle{thmstyle}
	{9pt}
	{9pt}
	{\itshape}
	{}
	{\bfseries\scshape}
	{.}
	{\newline}
	{}

\theoremstyle{thmstyle}

\begin{document}

\title{Sweetest taboo processes}

\author{Alain Mazzolo$^{1}$}
%%\email{alain.mazzolo@cea.fr}
%%\thanks{FAX: +33-1-6908-9490}

%%\affiliation{Den-Service d'\'etudes des r\'eacteurs et de math\'ematiques appliqu\'ees (SERMA), CEA, Universit\'e Paris-Saclay, F-91191 Gif-sur-Yvette, France }

\date{$^{1}$Den-Service d'\'etudes des r\'eacteurs et de math\'ematiques appliqu\'ees (SERMA), CEA, Universit\'e Paris-Saclay, F-91191 Gif-sur-Yvette, France}

\maketitle

\maketitle

\begin{abstract}
Brownian dynamics play a key role in understanding the diffusive transport of micro particles in a bounded environment. In geometries containing confining walls, physical laws determine the behavior of the random trajectories at the boundaries. For impenetrable walls, imposing reflecting boundary conditions to the Brownian particles leads to dynamics described by reflecting stochastic differential equations. In practice, these stochastic differential equations as well as their refinements are quite challenging to handle, and more importantly, many physical processes are better modeled by processes conditioned to stay in a prescribed bounded region. In the mathematical literature, these processes are known as taboo processes, and despite their simplicity, at least compared to the reflecting stochastic differential equations approach, are surprisingly not much exploited in physics. This paper explores some aspect of taboo processes and other constrained processes in simple geometries: Interval in one dimension, circular annulus in two dimensions, hollow sphere in three dimensions, and more. In particular, for the two-dimensional taboo process in a circular annulus, the Gaussian behavior of the stochastic angle is established.
\end{abstract}

%\maketitle
%%\noindent{\it Keywords}: Stochastic particle dynamics (theory), Brownian motion, Diffusion

%%\pacs{02.50.Ey, 05.10.Gg, 05.40.Jc}

%%\keywords{taboo process, constraint, stochastic differential equation, spectral gap}

%%%%%%%%%%%%%%%%%%%%%%%%%%%%%%%%%%%%%%%%%%%%%%%%%%%%%%%%%%%%%%%%%%%%%%%%%%%%%%%%%%%%%%%%%%%%%%%%%%%%%%%%%%%%%%%%
%%																						%%
%%                               INTRODUCTION                                                                 %%
%%																						%%
%%%%%%%%%%%%%%%%%%%%%%%%%%%%%%%%%%%%%%%%%%%%%%%%%%%%%%%%%%%%%%%%%%%%%%%%%%%%%%%%%%%%%%%%%%%%%%%%%%%%%%%%%%%%%%%%
\section{Introduction}
\label{sec_intro}

Brownian motion is a central object for understanding the diffusive transport of particles or the behavior of various living organisms (animals, insects, bacteria) and a vast literature ranging from neutrons transport~\cite{ref_book_Pazsit} to population dynamics~\cite{ref_Holmes} is dedicated to the subject. Such Brownian systems in confined spaces are generally highly sensitive to boundary conditions. Depending of the physical process, these boundary conditions can be of several kinds. For instance, in reactor physics, neutrons can either be absorbed on the surface of an absorbing medium or bounced on the surface of a reflector. The situation is even richer in biology and biophysics where a fraction of the individuals reaching the boundary can be reflected and the remainder is absorbed or passes through a partially permeable wall (a bio-membrane for example~\cite{ref_Fagan}). In this article, however, we will focus on impenetrable walls only, from a theoretical point of view, but also for simulations. This second aspect, that of the computer simulations of diffusion with reflection at the boundary of a domain, turns out to be unexpectedly complicated~\cite{ref_book_Schuss_1}. 
%%If the individuals that we to model were particles obeying classical dynamics, the situation on the walls would be rather simple: either the particle bounces with a reflected angle equal to that incident or it is reflected with an arbitrary secular angle (which depends on physics). 
Indeed, for a Brownian particle (and more generally for a diffusive particle), due to the absence of velocity, the situation at the boundary requires special care and one must add an extra ingredient when the particle touches the wall. Once on the wall, the particle must instantly feel its effect and immediately pushes back into the interior of the domain. Starting from a free diffusive process with drift term $\mu(x,t)$ and dispersion matrix $\sigma(x,t)$, the reflecting stochastic process $X_t$ in a bounded domain $\mathcal{D}$ is driven by a reflecting stochastic differential equation (see monograph~\cite{ref_Pilipenko} for an introduction),
\begin{equation}
\label{SDE_reflecting}
	dX_t = \mu(X_t,t) dt + \sigma(X_t,t) dW_t + \nu(X_t)dL_t ,
\end{equation}
\noindent where $W_t$ is a standard n-dimensional Brownian process (Wiener process), $\nu$ an inward unit normal vector field on the boundary $\partial D$ and $L_t$ is the so-called boundary local time of $X_t$. This last term, the local time, is understood as a measure of time the particle remains in the vicinity of the surface. More formally, a way of defining this (increasing) process is the following~\cite{ref_Zhou,ref_book_Karlin},
\begin{equation}
	L_t \equiv  \lim_{\epsilon \to 0} \frac{\int_0^t I_{D_{\epsilon}}(X_s) ds}{\epsilon}
\end{equation}
\noindent where $I_{A}$ is the indicator function of the set $A$ and $D_{\epsilon}$ is a strip region of width $\epsilon$ inside $\mathcal{D}$ containing $\partial D$. Others definitions of the local time involve the Dirac delta distribution~\cite{ref_book_Karlin,ref_Majumdar_Comtet}. Thus, defining the reflection process inside the boundary requires two unknown processes, $X_t$ and the local time process $L_t$. Extension of this theory to semi-permeable barriers takes the name of partially-reflected Brownian motion and has been recently achieved at the cost of introducing an additional exponentially distributed random variable $\chi$
%% of parameter $\Lambda > 0$ 
%%($ P\{\chi \ge x \} = \exp(-x/ \Lambda)$)
. Then the process is conditioned to stop when the stopping time 
%% $T_{\Lambda} = \inf \{t>0 : L_t \ge \chi\}$ 
of the boundary local time $L_t$ exceeds the random variable $\chi$ (this condition can be understood as an absorption after multiple reflections on the boundary of the domain). For a recent review on most aspects of partially-reflected Brownian motion see~\cite{ref_Grebenkov_reflected}. 
The analysis of partially-reflected stochastic processes needs subtle mathematical refinements (spread harmonic measure, Dirichlet-to-Neumann operator). We will not push forward our description of reflecting stochastic differential equations. The few preceding lines were intended to show that solving Eq.(\ref{SDE_reflecting}), or its generalizations,  even in simple geometries is a quite challenging task (a remark that remains valid for simulations~\cite{ref_Slominski}). Besides the inherent difficulties of this approach, mere reflections do not describe correctly many physical processes. For instance, as approaching an electrified barrier, ions feel progressively its effect. Identically, walls have a repulsive effect on (isolated) fishes, as confirmed by observations~\cite{ref_Gautrais}. In fact, this situation arises as soon as the species or the particle receive information from the wall (visual, smell, force), see Schuss's book where many examples from physics, chemistry and biology are presented~\cite{ref_book_Schuss_1}. Such processes do not touch the wall and are confined inside the domain in a way that is different from reflections on edges. Formally, these processes are constrained to stay forever inside the domain and in the mathematical literature, these processes are known under the name of taboo processes~\cite{ref_Knight,ref_Pinsky}. Although widely studied in the applied math literature and commonly used in financial fields~\cite{ref_Ikpe,ref_Li,ref_Abhyankar}, taboo processes, apart from the very recent works of Garbaczewski~\cite{ref_Garbaczewski,ref_Garbaczewski_2018} and that of Adorisio and coworkers~\cite{ref_Adorisio}, have not spread in the field of statistical physics yet. The purpose of this article is to remedy this situation by studying diverse examples of taboo processes in one, two and three dimensions, as well as in higher dimensions. The article also introduces and studies other families of processes forced to stay in a bounded region, in particular asymmetric processes.\\
%%cockroach, thanks to its antenna, can avoid obstacles without bump. 
%% if not intuitive, the process, which realizes this action, is well known and widely used, both from a theoretical point of view and for simulations. This is the so-called local time defined as follows:

The paper is organized as follows: in section~\ref{sec_Taboo1D}, after briefly reviewing the formalism of taboo processes, we focus on the important one-dimensional case, namely that of a taboo process living in an interval. We then extend this process to a more general class of taboo processes (including the sweetest one) and we also introduce a new kind of asymmetric simple one-dimensional confined process in section~\ref{sec_asymmetric_diffusion}. Next, in section~\ref{sec_Taboo2D}, the two-dimensional taboo process in a circular annulus is studied. In doing so, we will establish the Gaussian behavior of the angle of the taboo process in such a geometry. The three-dimensional case as well as higher dimensional taboo processes are then examined in section~\ref{sec_Taboo_higher_dimensions}. Finally, section~\ref{sec_Conclusion} contains concluding remarks. Monte Carlo simulations, also presented, support our theoretical findings in all dimensions. 

%%%%%%%%%%%%%%%%%%%%%%%%%%%%%%%%%%%%%%%%%%%%%%%%%%%%%%%%%%%%%%%%%%%%%%%%%%%%%%%%%%%%%%%%%%%%%%%%%%%%%%%%%%%%%%%%
%%																						%%
%%                 TABOO PROCESS                                                                              %%
%%																						%%
%%%%%%%%%%%%%%%%%%%%%%%%%%%%%%%%%%%%%%%%%%%%%%%%%%%%%%%%%%%%%%%%%%%%%%%%%%%%%%%%%%%%%%%%%%%%%%%%%%%%%%%%%%%%%%%%
\section{Taboo processes}
\label{sec_Taboo1D}
\subsection{General setting}

Taboo processes are usually defined as diffusion processes conditioned to remain forever in a bounded domain and we will precisely see what it means shortly. Before, we briefly recall their recent history. 
First of all, and to the best of our knowledge, in the context of probability theory the word "taboo" appears for the first time in Chung's book, published in 1960~\cite{ref_book_Chung}. However, the true beginning of the "taboos" as stochastic processes comes in 1969 with Knight's article~\cite{ref_Knight} who studies the particular case of a Brownian motion forced to remain in a finite interval $[0,a]$ or a semi-infinite domain $[a,+\infty)$ (Knight therefore derived the one-dimensional taboo process). To make a long story short, then in the mid-1980s, Pinsky studies the general case of diffusion processes conditioned to stay in an arbitrary bounded domain of $R^n$~\cite{ref_Pinsky}. Current development of taboo processes mainly concern L\'evy processes conditioned to stay in a bounded domain~\cite{ref_Garbaczewski_2018,ref_Lambert} and branching Brownian motions in a strip~\cite{ref_Harris}. Taboo processes find applications in mathematical finance where market information is often modeled by conditioning~\cite{ref_Ikpe,ref_Li,ref_Abhyankar}. It is only very recently that these processes have attracted, at last, the physicists' attention~\cite{ref_Garbaczewski,ref_Garbaczewski_2018,ref_Adorisio}.\\
 
Now, let us be more specific about what we mean by taboo processes. In the beginning of this paragraph, we specified that they were defined as diffusion processes conditioned to stay forever in a bounded domain $\mathcal{D}$. Taboo processes therefore belong to the family of constrained processes with respect to an event of zero probability, since for a sufficiently long time, the Brownian motion leaves a bounded domain with probability one. As we mentioned in two recent articles, conditioning a subtle object like a Brownian motion on events of zero probability is not an harmless task~\cite{ref_Mazzolo_Jstat,ref_Mazzolo_JMP}. 
Basically, there are two ways to achieve this conditioning. The first method consists in approximating the Brownian motion by a series of processes approaching it (typically by random walks) while the second method consists in approximating the conditioning event. The latter is known as the Doob's {\it{h}}-transform~\cite{ref_book_Doob}. A clear presentation of this technique is provided in chapter 15 of the book of Karlin and Taylor~\cite{ref_book_Karlin}. This approach is also outlined, from a physicist point of view, in the recent article~\cite{ref_Majumdar_Orland}. Since Knight and Pinsky's results on taboo processes were obtained thanks to an extension of Doob's {\it{h}}-transform, we recall its main feature. To this aim, let us consider an $n$-dimensional diffusion process $\{\bm{X}_t, 0 \le t \le T\}$ characterized by its drift $\bm{\mu}(\bm{x})$ and variance $\bm{\sigma}^2(\bm{x})$ coefficients. This diffusion satisfies the stochastic differential equation,
\begin{equation}
\label{diffusion_SDE}
	d \bm{X}_t  = \bm{\mu}(\bm{X}_t) \, dt +  \bm{\sigma}(\bm{X}_t) d\bm{W}_t \, ,
\end{equation}

\noindent where $\bm{\mu}(\bm{x})$ is a $n$-vector,  $\bm{\sigma}(\bm{x})$ is a $n \times n$ matrix, and $\bm{W}_t$ is a $n$-vector of standard Brownian motions (Wiener processes). We also introduce the generator $\mathcal{L}$ of the process,
\begin{equation}
\label{generator}
	\mathcal{L} f =  \sum_{i=1}^n  \mu_i(\bm{x}) \frac{\partial f}{\partial x_i}  + \frac{1}{2} \sum_{i=1}^n \sum_{j=1}^n \left( \bm{\sigma}(\bm{x}) \bm{\sigma}(\bm{x})^\mathsf{T} \right)_{i,j} \frac{\partial^2 f}{\partial x_i \partial x_j} \, , 
\end{equation}

\noindent and 

\begin{equation}
\label{generator_adjoint}
	 \mathcal{L}^{\dagger} f  = - \sum_{i=1}^n \frac{\partial  (\mu_i(\bm{x}) f)}{\partial x_i} + \frac{1}{2}  \sum_{i=1}^n \sum_{j=1}^n \frac{\partial^2 \left(   \left( \bm{\sigma}(\bm{x}) \bm{\sigma}(\bm{x})^\mathsf{T} \right)_{i,j} f \right)}{\partial x_i \partial x_j} 
\end{equation}

\noindent its adjoint~\cite{ref_book_DelMoral}. Next, we consider an event $G(T)$ between two times 0 and $T$. 
To fix the ideas, this event can be the one where the process is constrained to stay in the domain until time $T$, in this case $G(T)= \{\{\bm{X}_t\}_{0 \le t \le T} \in \mathcal{D} \}$. Now, let $\{\bm{X}^*_t, 0 \le t \le T\}$ be the process conditioned on its state at time $T$. Then, the variance $\bm{\sigma}^{*2}(\bm{x},t)$ and the drift $\bm{\mu}^*(\bm{x},t)$ of the constrained process are respectively given by:

\begin{equation}
\label{constrained_drift_variance}
  \left\{
      \begin{aligned}
        & \bm{\sigma}^{*}(\bm{x},t) = \bm{\sigma}(\bm{x})   \\
        & \bm{\mu}^*(\bm{x},t) = \bm{\mu}(\bm{x})  + \bm{\sigma}(\bm{x})\bm{\sigma}(\bm{x})^\mathsf{T}   \frac{\bm{ \nabla} \pi(\bm{x},t; G(T))}{\pi(\bm{x},t; G(T))} \, , \\
      \end{aligned}
    \right.
\end{equation}

\noindent where $\pi(\bm{x},t; G(T))$ is the probability that from the state value $\bm{x}$ at time $t$, the sample path of $\bm{X}_t$ satisfies the desired constraint $G(T)$ at time $T$. Consequently, Doob's method requires the calculation of an often-intricate probability and its analytic expression is unfortunately seldom known. When this quantity is available, Doob's technique have been successfully applied to various kind of conditioned processes~\cite{ref_book_Karlin,ref_Majumdar_Orland,ref_Orland,ref_Szavits,ref_Baudoin,ref_Chetrite}. When the closed-form of the probability is not available several techniques have been developed, including constraint in abstract Wiener space~\cite{ref_Mazzolo_Jstat,ref_Sottinen}, stochastic control techniques~\cite{ref_Mazzolo_JMP,ref_Pavon}, and series expansion~\cite{ref_Gorgens}. Moreover, when the process is forced to stay inside the domain until time $T$, Pinsky established the following strong relationship~\cite{ref_Pinsky},

\begin{equation}
 	\lim_{T \to \infty} \frac{\bm{ \nabla} \pi(\bm{x},t; G(T))}{\pi(\bm{x},t; G(T))} = \frac{\bm{ \nabla} \varphi_1(\bm{x})}{\varphi_1(\bm{x})} \, ,
\end{equation}

\noindent where $\varphi_1(\bm{x})$ is the first eigenfunction of the operator $-\mathcal{L}$ on the domain $\mathcal{D}$,  with Dirichlet boundary conditions. We give a simple proof of this result in appendix~\ref{appendix_1}. From the preceding expression, one can understand in a heuristic way that the process will remain in $\mathcal{D}$ forever. Indeed, since $\varphi_1(\bm{x})$ vanishes on the boundary of the domain, when the process approaches the boundary the drift term explodes, forcing the process to stay inside the domain. In addition, another result obtained by Pinsky states that the invariant density $\Psi(x)$~\footnote{We postpone the proof of the existence of an invariant density for the taboo process in section~\ref{subsec_invariant_density} where it will be established in the general setting of generalized taboo processes.}(the long time behavior of the process as $t \to \infty$) is given by~\cite{ref_book_Karlin}

\begin{equation}
\label{Invariant_density_Pinsky}
	 \Psi(x) = \varphi_1(\bm{x}) \varphi_1^{\dagger}(\bm{x}) \, ,
\end{equation}

\noindent where $\varphi_1^{\dagger}(\bm{x})$ is the first eigenfunction of the adjoint operator $-\mathcal{L}^{\dagger}$ on the domain $\mathcal{D}$ with Dirichlet boundary conditions, provided that $\varphi_1(\bm{x})$ and $\varphi_1^{\dagger}(\bm{x})$ are normalized ($\int_{\mathcal{D}} \varphi_1(\bm{x}) \varphi_1^{\dagger}(\bm{x}) d\bm{x}= 1$). To summarize, the process $\bm{X}^*_t$ constrained to remain forever in the domain satisfies the stochastic differential equation \\
\begin{equation}
\label{taboo_SDE}
	d \bm{X}^*_t  = \left[ \bm{\mu}(\bm{\bm{X}^*_t})  + \bm{\sigma}(\bm{\bm{X}^*_t})\bm{\sigma}(\bm{\bm{X}^*_t})^\mathsf{T} \frac{\bm{ \nabla} \varphi_1(\bm{\bm{X}^*_t})}{\varphi_1(\bm{\bm{X}^*_t})}  \right] \, dt +  \bm{\sigma}(\bm{X}^*_t) d\bm{W}_t \, .
\end{equation}

\noindent This equation calls for at least two remarks. First of all, if we apply the method of conditioning to a domain $\mathcal{D'} \subset \mathcal{D}$, the new process will remain in $\mathcal{D'}$ forever and thus also in $\mathcal{D}$. This remark shows that the taboo process is not the only process to be forced to stay inside $\mathcal{D}$. More importantly, let us consider a drift term of the form
\begin{equation}
\label{modified_drift}
 	 \alpha \times \left[ \bm{\mu}(\bm{x})  + \, \bm{\sigma}(\bm{x})\bm{\sigma}(\bm{x})^\mathsf{T}\frac{\bm{ \nabla} \varphi_1(\bm{x})}{\varphi_1(\bm{x})} \right] \, ,
\end{equation}
\noindent where $\alpha$ is a positive number. For any $\alpha \ge 1$, when this process approaches $\partial \mathcal{D}$, this new (inward) drift will be strong enough to keep the process in $\mathcal{D}$. In the one-dimensional case studied in the next paragraph, we will see that, quite surprisingly, this result remains true even for values of $1/2 \le \alpha \le 1$. So what distinguishes the taboo process (corresponding to $\alpha=1$) from this family of processes that does not leave $\mathcal{D}$? The response is given in the articles~\cite{ref_Banuelos_Pang,ref_Banuelos_Mendez} where it is established that the taboo process behaves similar to the reflected Brownian motion near the boundary. This opens beautiful perspectives to model various behaviors of diffusions near the walls. Indeed, by considering a drift term of the form~Eq.(\ref{modified_drift}), the parameter $\alpha$ will allow us to control the rigidity of the impenetrable barrier.  For $\alpha>1$, in some way, the walls become more repulsive than a standard impenetrable barrier whereas for certain values of $\alpha<1$ they become softer (while remaining impenetrable). In the remainder of this article, the smallest value of $\alpha$ for which the process remains confined within its boundaries will take the denomination of {\it{sweetest taboo process}}.

%%%%%%%%%%%%%%%%%%%%%%%%%%%%%%%%%%%%%%%%%%%%%%%%%%%%%%%%%%%%%%%%%%%%%
%%         one-dimensional TABOO PROCESS                           %%
%%%%%%%%%%%%%%%%%%%%%%%%%%%%%%%%%%%%%%%%%%%%%%%%%%%%%%%%%%%%%%%%%%%%%
\subsection{one-dimensional taboo process}
The simplest taboo process is the one evolving in one dimension, corresponding to a stochastic process confined within a fixed interval. This process has been extensively studied by Knigth~\cite{ref_Knight} and Pinsky~\cite{ref_Pinsky} and the purpose of this paragraph is (above all) to get familiar with this fundamental taboo process. 
First, we will recall the standard procedure to obtain the stochastic differential equation describing the taboo process. Then, we will derive the stationary distribution and analyze in detail the speed of convergence of the process towards this limit. We will also look at the average behavior of the taboo process as a function of time. Monte Carlo simulations will illustrate our points.\\
%%An interesting result concerns however the average behavior of the process. 

\noindent Let us consider a fixed interval $[-L,L]$ and a diffusion process $X_t$ (starting inside the interval, $X_0 \in [-L,L]$), with constant parameters $\mu$ and $\sigma$. This process corresponds to a Brownian motion with constant drift and satisfies the scalar stochastic differential equation,
\begin{equation}
\label{diffusion_SDE-scalar}
	d X_t  = \mu \, dt +   \sigma dW_t \, ,
\end{equation}
whose generator is 
\begin{equation}
\label{generator_scalar}
	\mathcal{L} f =    \mu  \frac{d f}{d x}  + \frac{\sigma^2}{2}  \frac{d^2 f}{d x^2} \, , 
\end{equation}
\noindent and
\begin{equation}
\label{generator_adjoint_scalar}
	\mathcal{L}^{\dagger} f =   - \mu  \frac{d f}{d x}  + \frac{\sigma^2}{2} \frac{d^2 f}{d x^2} \, 
\end{equation}
\noindent its adjoint. With these parameters, the corresponding (scalar) taboo stochastic differential equation Eq.(\ref{taboo_SDE}) reduces to
\begin{equation}
\label{taboo_SDE_1D}
	d X^*_t  = \left[ \mu + \sigma^2 \frac{\varphi_1'(X^*_t)}{\varphi_1(X^*_t)}  \right]  dt + \sigma \, dW_t \, ,
\end{equation}
\noindent where $\varphi_1(x)$ is the first eigenfunction of the operator $-\mathcal{L}$ on the domain $[-L,L]$, with Dirichlet boundary conditions. Let $\lambda_1$ be its eigenvalue. The equation $-\mathcal{L}\varphi_1(x) = \lambda_1 \varphi_1(x)$ with Dirichlet boundary conditions writes 

\begin{equation}
\label{eq_first_eigenfunction_1D}
  \left\{
      \begin{aligned}
	  &\frac{\sigma^2}{2} \frac{d^2 \varphi_1(x)}{d x^2} + \mu  \frac{d \varphi_1(x)}{d x} = -\lambda_1 \varphi_1(x)
       \\
       & \varphi_1(-L) = \varphi_1(L) = 0 \, .\\
      \end{aligned}
    \right.
\end{equation}

\noindent Its solution is given by
\begin{equation}
\label{phi1-1-dimension}
	 \varphi_1(x) = e^{-\mu x/\sigma^2} \cos \left( \frac{\pi x}{2L} \right), 
\end{equation}
\noindent and the lowest eigenvalue is~\cite{ref_book_Redner}
\begin{equation}
\label{lambda1-1-dimension}
	 \lambda_1 = \frac{\mu^2}{2 \sigma^2} + \frac{\pi^2 \sigma^2}{8 L^2} .
\end{equation}
\noindent Inserting the logarithmic derivative of $\varphi_1(x)$ in Eq.(\ref{taboo_SDE_1D})
 leads immediately to the stochastic differential equation satisfied by the taboo process $X_t^*$ within the segment $[-L,L]$,
\begin{equation}
\label{taboo_SDE_1D_interval}
  \left\{
      \begin{aligned}
        dX^*_t & = - \frac{\pi \sigma^2}{2L} \tan \left( \frac{\pi X^*_t}{2L}   \right)  dt  +  \sigma \, dW_t %% \qquad  0 \leq t < T 
       \\
        X^*_0 & = x_0 \in [-L,L] , \\
      \end{aligned}
    \right.
\end{equation}
\noindent which is independent of $\mu$. Due to the tangent function, as expected the drift term of the taboo process diverges (with the appropriate sign) when $X^*_t$ approaches the boundaries. Note also, that the drift term is equal to zero when $X^*_t = 0$, corresponding to the middle of the interval $[-L,L]$, meaning that the taboo process behaves as a pure Brownian motion at equal distance from the boundaries. \\

\noindent In order to fully characterize the taboo process, we give the analytical form of the stationary distribution and closely examine the speed of convergence of the process towards its equilibrium density. Following Pinsky's approach~\cite{ref_Pinsky}, the long-time limit $\Psi(x)$ of the taboo process is given by Eq.(\ref{Invariant_density_Pinsky}) and required also the first eigenfunction $\varphi^{\dagger}_1(x)$ of the adjoint operator $-\mathcal{L}^{\dagger}$ on the domain $[-L,L]$, with Dirichlet boundary conditions, $\varphi^{\dagger}_1(-L)= \varphi^{\dagger}_1(L)=0$. Proceeding in the same way as for the eigenfunction $\varphi_1(x)$, we get that $\varphi^{\dagger}_1(x) = C e^{\mu x/\sigma^2} \cos \left( \frac{\pi x}{2L} \right)$. Similarly its eigenvalue $\lambda^{\dagger}_1$ is~\footnote{The fact that the lowest eigenvalue and the adjoint lowest eigenvalue are the same is highlighted in appendix~\ref{appendix_2} in the general setting where $\mu$ and $\sigma$ are not constant.}
\begin{equation}
\label{lambda1-dagger-1-dimension}
	 \lambda^{\dagger}_1 = \frac{\mu^2}{2 \sigma^2} + \frac{\pi^2 \sigma^2}{8 L^2} = \lambda_1.
\end{equation}
\noindent The constant $C$ is obtained thanks to the normalization $\int_{-L}^L \varphi_1(x) \varphi^{\dagger}_1(x) dx = 1$ and finally,
\begin{equation}
\label{phi1-dagger-1-dimension}
	 \varphi^{\dagger}_1(x) = \frac{1}{L} e^{\mu x/\sigma^2} \cos \left( \frac{\pi x}{2L} \right).
\end{equation}
\noindent Inserting the expressions of the eigenfunction Eq.(\ref{phi1-1-dimension}) and its adjoint Eq.(\ref{phi1-dagger-1-dimension}) into Eq.(\ref{Invariant_density_Pinsky}) gives the long time behavior of the one-dimensional taboo process
\begin{equation}
\label{invariant_measure_1_dimension}
	 \Psi(x) = \frac{1}{L} \cos^2 \left( \frac{\pi x}{2L} \right) ,
\end{equation}
\noindent which is, as noted by Pinsky~\cite{ref_Pinsky}, independent of the original drift $\mu$. Note that the invariant stationary density can be obtained by standard tools from stochastic calculus, without resorting to the lowest eigenfunction, as done in appendix~\ref{appendix_3}. The preceding expression gives the invariant density but it remains to determine how fast the taboo process converges to its asymptotic shape?\\
The answer to this important question is provided by an article by Smits~\cite{ref_Smits}, where the author establishes that the taboo process converges towards its equilibrium limit exponentially quickly and that the rate of convergence is related to the difference between the two lowest eigenvalues of the operator $-L$ on the domain, with Dirichlet boundary conditions. The difference between the first two eigenvalues $\lambda_2 - \lambda_1$ is often called the (Dirichlet) spectral gap (or fundamental gap). When $\mathcal{L}$ is the Laplace operator, the recent review by Grebenkov and Nguyen~\cite{ref_Grebenkov_laplacian} gives comprehensive accounts of the properties of eigenvalues and eigenfunctions of the Laplacian in bounded domains, with various boundary conditions. For a taboo process evolving in an interval, the eigenvalue equation with Dirichlet boundary conditions writes,

\begin{equation}
\label{eq_eigenfunction_1D}
  \left\{
      \begin{aligned}
	  &\frac{\sigma^2}{2} \frac{d^2 \varphi_n(x)}{d x^2} + \mu  \frac{d \varphi_n(x)}{d x} = -\lambda_n \varphi_n(x)
       \\
       & \varphi_n(-L) = \varphi_n(L) = 0 \, .\\
      \end{aligned}
    \right.
\end{equation}

\noindent The eigenvalues, easily found, are, 
\begin{equation}
\label{lambda_n-1-dimension}
	 \lambda_n = \frac{\mu^2}{2 \sigma^2} + \frac{\pi^2 \sigma^2}{8 L^2} n^2 \, ,
\end{equation}
\noindent and the spectral gap is
\begin{equation}
\label{spectral_gap_1D}
 	\lambda_2 - \lambda_1 = 3 \frac{\pi^2 \sigma^2}{8 L^2} .
\end{equation}

\begin{figure}[h]
\centering
\includegraphics[width=3.in,height=3.5in,angle=-90]{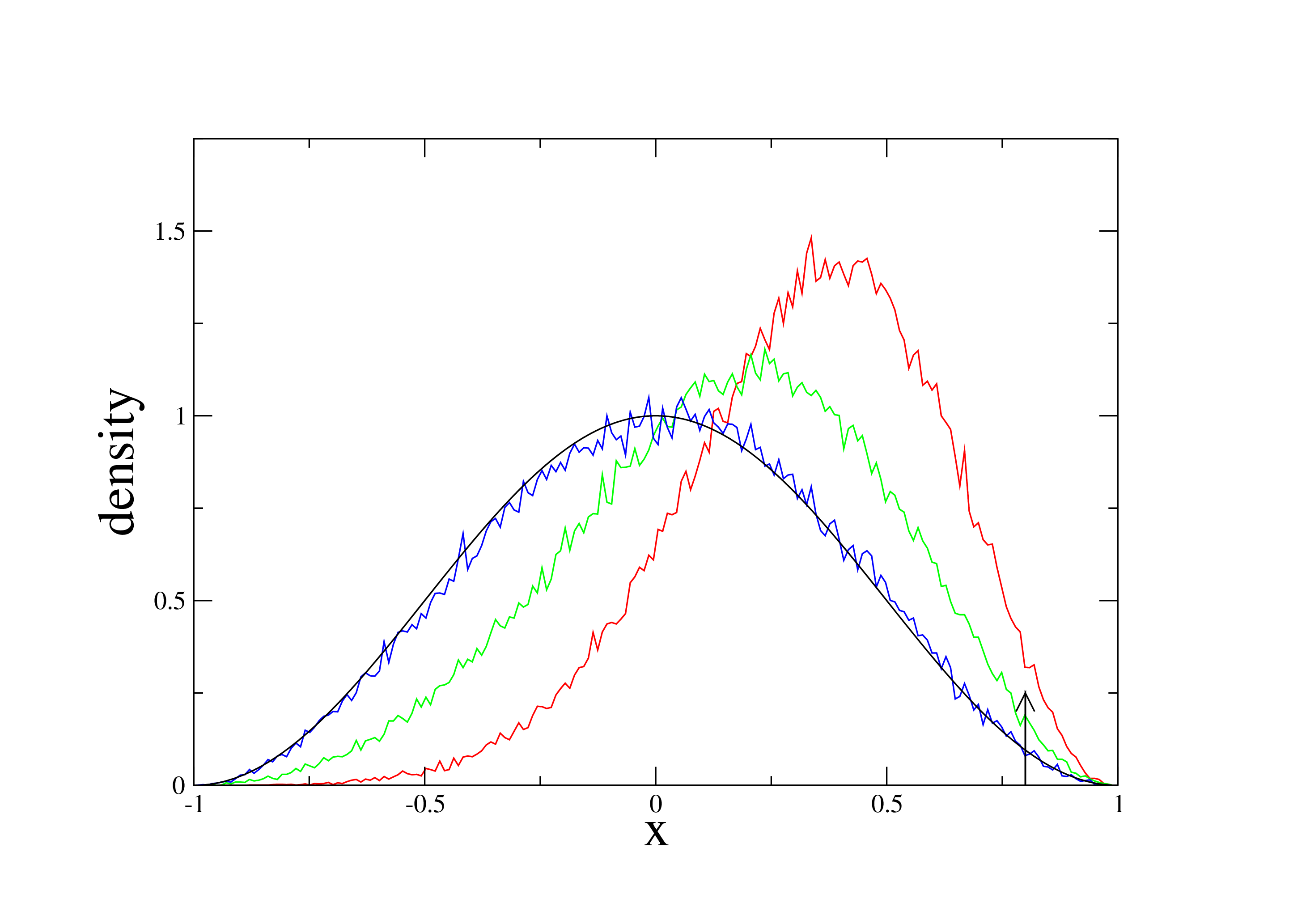}
\setlength{\abovecaptionskip}{15pt} 
\caption{Evolution of the density of the one-dimensional taboo process within the interval $[-1,1]$: the red line corresponds to $t = 0.2$, the green one to $t = 0.4$ and the blue one to $t = 1$. All curves start at $x_0 = 0.8$ (symbolized by a black arrow) and are obtained from simulations of $10^5$ trajectories with time step $10^{-2}$. The black curve is the theoretical profile of the stationary density as given by Eq.(\ref{invariant_measure_1_dimension}). For $t = 1 $ the taboo process has very well converged towards its stationary density.}
\label{fig1}
\end{figure}

\noindent The rate of convergence is given by $\sim \exp(-(\lambda_2 - \lambda_1) t)$~\cite{ref_Smits}. More precisely, in the limit of long times, the transition density $p(x,t|y,0)$ and the stationary density $\Psi(x)$ satisfy the general formula

\begin{equation}
	\left| p(x,t|y,0) - \Psi(x) \right| \le K e^{-(\lambda_2 - \lambda_1)t} \Psi(x) ,
\end{equation}

\noindent where $K$ is a suitable positive constant. We thus have an exponentially fast convergence to equilibrium as stated in references~\cite{ref_Garbaczewski,ref_Garbaczewski_2018} where the detailed expression of the transition probability density function of the taboo process is also given. For standard values of the diffusion coefficient, say $\sigma = 1$, and a typical size of the domain $L = 1$, the taboo process will have converged to its stationary density for times greater than $\tau = 1/(\lambda_2 - \lambda_1) \sim 0.27$.
%%which is relatively small. 
Figure~\ref{fig1}
plots the theoretical invariant density as well as three densities at different times. The simulations show that for a simulation time of $1 \gg \tau \sim 0.27$, the density has already converged very well towards its asymptotic shape. \\ 

\subsection{one-dimensional sweetest taboo process}
\label{sub_sec_sweetest_1D}
\subsubsection{Definition and boundary classification}
In the previous paragraph, we described the taboo processes evolving in confined environments with strictly rigid walls. However, as we have mentioned, many natural processes evolve in geometries bounded with more or less impenetrable borders. In order to model these different kinds of walls, we consider an extension of the previous taboo process by introducing a positive parameter $\alpha$ that multiplies the original taboo drift term. We call this new process the {\it{generalized taboo process}}\footnote{In the mathematical literature where $\sigma$ often takes the value $1$, the generalized taboo process is also denominated Legendre process (see~\cite{ref_book_Revuz_Yor}, p. 357) or radial affine Dunkl process~\cite{ref_Chapon}. The generalized taboo process is also considered as a particular Bessel-like process in the study of radial Schramm-Loewner evolution~\cite{ref_book_Lawler}.}. For $\alpha > 1$, the drift term is greater than the drift of the taboo process and the walls become more repulsive. For $\alpha = 0$, the process corresponds to the original free diffusion and the wall are transparent. Thus, there exists a minimal value $\alpha_s\in ]0,1]$ for which the walls remain impenetrable. As we stated previously, we call {\it{sweetest taboo process}} the stochastic process associated with the value $\alpha_s$.\\

\noindent The generalized taboo process is described by the stochastic differential equation,
\begin{equation}
\label{modified_taboo_SDE_1D_interval}
  \left\{
      \begin{aligned}
        dZ_t & = - \alpha \frac{\pi \sigma^2}{2L} \tan \left( \frac{\pi Z_t }{2L}  \right)  dt  +  \sigma \, dW_t %% \qquad  0 \leq t < T 
       \\
        Z_0 & = x_0 \in [-L,L] \, . \\
      \end{aligned}
    \right.
\end{equation}
\noindent To understand the behavior of the process near its borders, we must determine the values of $\alpha$ for which the process remains confined in the interval $[-L,L]$. It is thus a question of studying the behavior of the process at its left boundary $-L$, and at its right boundary $L$.  This can be achieved thanks to the boundary classification of diffusion processes, as exposed in the book of Karlin and Taylor~\cite{ref_book_Karlin}. Since we are interested only in processes confined in their state space (here the interval $[-L,L]$), we need the sole definition of an entrance boundary. This one is defined as a boundary that cannot be reached from the interior of the state space (where the process starts). Consequently, in order to obtain a process constrained to remain forever in the interval $[-L,L]$, the parameter $\alpha$ 
must be chosen in such a way that the left and the right boundaries are entrance boundaries. We focus on the left boundary, i.e. when $Z_t$ approaches $-L$ (a similar reasoning applies to the right boundary when the process is near $L$). Criteria for a border to be an entrance boundary are summarized in Eq.(\ref{theorem_entrance}) of appendix~\ref{appendix_3}. Let us consider a one-dimensional diffusion process with drift $\mu(x)$ and variance $\sigma^2(x)$, we introduce the function s(x) (see appendix~\ref{appendix_3} for more details regarding these functions),

\begin{equation}
\label{s_function_generalized_taboo_1D_interval}
	s(x) = \exp \left(- \int_{0}^{x} \frac{2 \mu(\xi)}{\sigma^2(\xi)} d\xi \right) 
\end{equation}

\noindent as well as the scale measure $S(l,x]$, the speed measure $M(l,x]$,
\begin{equation}
\label{speed_measure}
        S(l,x]  =  \int_{l}^x s(\eta)  d\eta  \, , \qquad M(l,x] =  \int_{l}^x \frac{d\eta} {\sigma^2(\eta) s(\eta)} 
\end{equation}
\begin{equation}
\label{speed_measure_N}
         \mathrm{and}  \qquad N(l)  =  \int_{l}^x M(l,\xi]  s(\xi) d\xi \, .
\end{equation}

\noindent Then, according to Eq.(\ref{theorem_entrance}) of appendix~\ref{appendix_3}: $-L$ is an entrance boundary if $S(-L,x] = \infty$ while $N(-L) < \infty$.
%%it is sufficient to check that $S(0,x] = \infty$ while $N(0) < \infty$. All functions are defined in this appendix. 
With the parameters of the generalized taboo process $\mu(x)= - \alpha \pi \sigma^2/(2L) \tan \left( \pi x/(2L)  \right)$ and $\sigma(x) =  \sigma$, we get successively,
\begin{equation}
\label{s_function_generalized_taboo_1D_interval_def}
        s(x) = \exp \left(\int_{0}^{x} \alpha \frac{\pi \sigma^2}{2L} \tan \left( \frac{\pi}{2L} \xi  \right) d\xi \right) = \frac{1}{\left[\cos\left(\frac{\pi x}{2L} \right) \right]^{2 \alpha} } \, ,
\end{equation}
\noindent and
\begin{equation}
\label{scale_function_generalized_taboo_1D_interval}
    S(-L,x] =  \int_{-L}^x s(\eta)  d\eta = \int_{-L}^x \frac{d\eta}{\left[\cos\left(\frac{\pi \eta}{2L} \right) \right]^{2 \alpha} } \, . \\    
\end{equation}

%%\noindent Since $\cos(\frac{\pi x}{2L}) \underset{x \to -L}{\sim\,} \frac{\pi (x+L)}{2L} $, 
\noindent Since $\cos(\frac{\pi x}{2L}) \underset{x \to -L}{\sim\,} \frac{\pi (x+L)}{2L} $, the preceding integral diverges for $\alpha \ge 1/2$. It remains to verify that $N(-L)$ is finite. Combining Eqs.(\ref{speed_measure},\ref{speed_measure_N})  and Eq.(\ref{scale_function_generalized_taboo_1D_interval}) we get that,\\
%%\begin{equation}
%%    N(0) =  \int_0^x  \left( \int_0^{\xi} \frac{d\eta}{\sigma(\eta)^2 s(\eta)} \right) s(\xi) d\xi = \frac{1}{\sigma^2} \int_0^x \left( \int_0^{\xi} \left[\sin\left(\frac{\pi }{R}\eta \right) \right]^{\frac{2 \alpha}{\sigma^2}} d\eta \right) \left[\sin\left(\frac{\pi }{R}\xi \right) \right]^{\frac{2 \alpha}{\sigma^2}} d\xi \, .
%%\end{equation}
%%
%%\begin{equation}
%%    N(-L) =  \int_{-L}^x  \left( \int_{-L}^{\xi} \frac{d\eta}{\sigma(\eta)^2 s(\eta)} \right) s(\xi) d\xi = \frac{1}{\sigma^2} \int_{-L}^x \int_{-L}^{\xi} \left[\frac{ \cos\left(\frac{\pi \eta}{2L} \right)}{\cos\left(\frac{\pi \xi}{2L} \right)} \right]^{2 \alpha} d\eta \, d\xi \, + \frac{1}{\sigma^2} \int_{-L}^x \int_{-L}^{\xi} \left[\frac{ \cos\left(\frac{\pi \eta}{2L} \right)}{\cos\left(\frac{\pi \xi}{2L} \right)} \right]^{2 \alpha} d\eta \, d\xi \, .
%%\end{equation}
\begin{equation}      
	\begin{aligned}
    N(-L) =  \int_{-L}^x  \left( \int_{-L}^{\xi} \frac{d\eta}{\sigma(\eta)^2 s(\eta)} \right) s(\xi) d\xi & = \frac{1}{\sigma^2} \int_{-L}^0 \int_{-L}^{\xi} \left[\frac{ \cos\left(\frac{\pi \eta}{2L} \right)}{\cos\left(\frac{\pi \xi}{2L} \right)} \right]^{2 \alpha} d\eta \, d\xi \\
		& + \frac{1}{\sigma^2} \int_{0}^x \int_{-L}^{\xi} \left[\frac{ \cos\left(\frac{\pi \eta}{2L} \right)}{\cos\left(\frac{\pi \xi}{2L} \right)} \right]^{2 \alpha} d\eta \, d\xi \, .      
	\end{aligned}
\end{equation}

\noindent The second integral is always finite. Besides, for $\alpha \ge 1 / 2$ we have $ 0 \le  \left[\cos\left(\frac{\pi \eta}{2L} \right) / \cos\left(\frac{\pi \xi}{2L} \right)  \right]^{2 \alpha} \le 1 $ in the first integral and its integrand thus lies between $0$ and $1$, therefore $N(-L)  < \infty$. A similar argument applies to the right boundary $L$. We can conclude that both boundaries are entrance and hence that the stochastic process driven by the stochastic differential equation Eq.(\ref{modified_taboo_SDE_1D_interval}) with  $\alpha \ge 1 / 2$ remains forever inside the interval $[-L,L]$. For $\alpha < 1/2$ the walls become permeable, allowing the process to cross the boundaries\footnote{More precisely, if $-1 /2 < \alpha < 1 / 2$ both endpoints $-L$ and $L$ are regular boundaries (meaning that the process can both enter and leave from the boundaries). If $\alpha \le -1 /2$ endpoints are exit boundaries (meaning that once the process reaches (or starts form) such boundaries, it cannot (re)enter inside the domain. This can be demonstrated by arguments similar to those discussed in the paragraph \ref{sub_sec_sweetest_1D} and thanks to the boundary classification established by Karlin and Taylor~\cite{ref_book_Karlin}. One can also have a quick idea of these results by noting that for values of $Z_t$ close to $-L$, the shifted process $\tilde{Z}_t = Z_t +L$ (Eq.(\ref{modified_taboo_SDE_1D_interval})) behaves as $d\tilde{Z}_t \sim  (\alpha \sigma^2/\tilde{Z}_t) dt + \sigma dW_t$. This last equation corresponds to a Bessel process whose boundary classification is given in the article~\cite{ref_Martin}. We will not develop these points in the present article.}. In the following, we define the sweetest taboo process $S_t$ as the one corresponding to the smallest value $\alpha_s$ for which the boundaries remain entrance. Hence $\alpha_s = 1/2$, and the stochastic differential equation satisfied by the sweetest taboo process in the domain $[-L,L]$ is thus
\begin{equation}
\label{sweetest_taboo_SDE_1D_interval}
  \left\{
      \begin{aligned}
        dS_t & = - \frac{\pi \sigma^2}{4L} \tan \left( \frac{\pi S_t}{2L}   \right)   dt  +  \sigma \, dW_t  
       \\
        S_0 & = x_0 \in [-L,L] . \\
      \end{aligned}
    \right.
\end{equation}

%%%%%%%%%%%%%%%%%%%%%%%%%%%%%%%%%%%%%%%%%%%%%%%%%%%%%%%%%%%%%%%%%%%%%
%%                INVARIANT DENSITY                                %%
%%%%%%%%%%%%%%%%%%%%%%%%%%%%%%%%%%%%%%%%%%%%%%%%%%%%%%%%%%%%%%%%%%%%%
\subsubsection{Invariant density}
\label{subsec_invariant_density}
\begin{figure}[h]
\centering
\includegraphics[width=3.0in,height=3.5in,angle=-90]{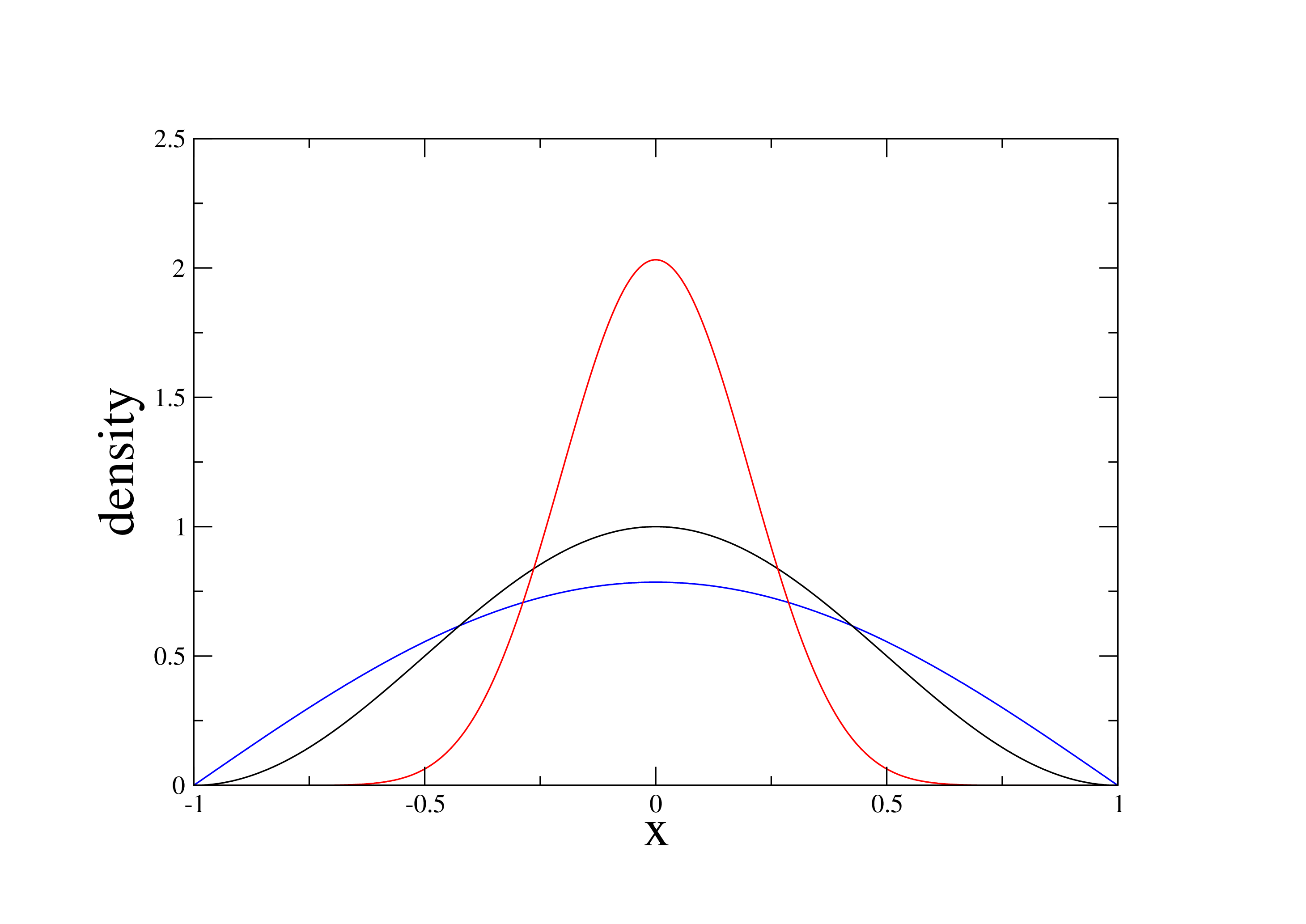}
\setlength{\abovecaptionskip}{15pt} 
\caption{Invariant density of the generalized taboo process in the interval $[-1,1]$ as given by Eq.(\ref{stationary_density_taboo_generalized}). Blue line: sweetest taboo process ($\alpha = 1 /2 $), black line: standard taboo process ($\alpha = 1$), red line: generalized taboo process with $\alpha = 5 $. When the parameter $\alpha$ increases, i.e. when the walls become more repulsive, the stationary distribution narrows around the middle of the interval where the process spends much of its time.}
\label{fig2}
\end{figure}

As noted in appendix~\ref{appendix_3} when both ends of a one-dimensional diffusion process are entrance boundaries, the process is ergodic, meaning that the distribution of the diffusion converges to a (unique) stationary distribution $\Psi(x)$ as $t \to \infty$. In this appendix, we also pointed out that this invariant distribution is related to the scale function and speed density through the formula
\begin{equation}
      \Psi(x) =  \frac{m(x)}{\int_{-L}^{L} m(\eta) d\eta} = \frac{1}{\sigma^2(x) s(x) \int_{-L}^{L} \frac{d\eta}{\sigma^2(\eta) s(\eta)} } \, .
\end{equation} 
\noindent Inserting the diffusion and drift coefficients of the generalized taboo process which are $\sigma(x)=\sigma$ and $\mu(x)= - \alpha \frac{\pi \sigma^2}{2L} \tan \left( \frac{\pi x}{2L}  \right)$  into the preceding equation leads, after a straightforward calculation, to (recall that $s(x)$ is given by Eq.(\ref{s_function_generalized_taboo_1D_interval}))
\begin{equation}
\label{stationary_density_taboo_generalized}
      \Psi(x) =  \frac{\sqrt{\pi}}{2L} \frac{\Gamma\left(1+ \alpha \right)}{\Gamma\left(\frac{1}{2} + \alpha \right)} \left( \cos\left(  \frac{\pi x}{2L} \right) \right)^{2 \alpha} \, ,
\end{equation}
\noindent where $\Gamma$ denotes the Euler Gamma function. When $\alpha = 1$, corresponding to the standard taboo process, we recover the invariant density of the taboo process $\Psi(x) =  1/L \cos^2 ( \pi x /2L ) $ (Eq.(\ref{invariant_measure_1_dimension})). When $\alpha =1/2 $, corresponding to the sweetest taboo process, the invariant density reduces to
\begin{equation}
\label{stationary_density_sweetest_taboo}
      \Psi_{s}(x) =  \frac{\pi}{4 L} \, \cos\left( \frac{ \pi x}{2L} \right)\, .
\end{equation}
\noindent As expected, since both endpoints act in the same way on the process, the invariant density Eq.(\ref{stationary_density_taboo_generalized}) is symmetric with respect to the middle of the interval. Moreover, when $x$ approaches $-L$ (again a similar reasoning applies to the right boundary) we have,
\begin{equation}
      \frac{d \Psi(x)}{dx} \underset{x \to -L}{\sim\,}  \left( 1+\frac{x}{L}  \right)^{2 \alpha-1} 
	                      \frac{\alpha  \pi^{2 \alpha + \frac{1}{2}} }{2^{2 \alpha} L^2} \frac{\Gamma\left(1+ \alpha \right)}{\Gamma\left(\frac{1}{2} + \alpha \right)} ,
\end{equation}
\noindent and $\Psi'(-L) \ne 0 $ for $\alpha = 1/2 $ only. The sweetest taboo process is the sole generalized taboo process for which the invariant density does not start softly at the boundaries. In others words, the repulsion from the walls is sweet enough to let the process wander near them. The shape of the stationary density profile is plotted on Fig.~\ref{fig2}. The maximum of the curve is $ \sqrt{\pi} \Gamma\left(1+ \alpha \right)/ (2L \, \Gamma\left(1/2 + \alpha \right))$ and tends to infinity as $\alpha \to \infty$, meaning that in the limit of infinitely repulsive walls, the process stays in the middle of the interval.

%%%%%%%%%%%%%%%%%%%%%%%%%%%%%%%%%%%%%%%%%%%%%%%%%%%%%%%%%%%%%%%%%%%%%
%%                  RATE OF CONVERGENCE                            %%
%%%%%%%%%%%%%%%%%%%%%%%%%%%%%%%%%%%%%%%%%%%%%%%%%%%%%%%%%%%%%%%%%%%%%
\subsubsection{Rate of convergence}
\label{subsec_rate_of_convergence}
In order to characterize the generalized taboo process, besides the invariant density, the knowledge of the speed of convergence to the stationary density is also crucial. As we already mentioned, the rate of convergence is given by the spectral gap of the $-\mathcal{L}$ operator, i.e.  the difference between the two lowest eigenvalues. For the generalized taboo process the infinitesimal operator is $\mathcal{L_T}  =   \frac{\sigma^2}{2}  \frac{d^2 }{d x^2} - \alpha \frac{\pi \sigma^2}{2L} \tan \left( \frac{\pi x}{2L} \right)  \frac{d }{dx} $ and the eigenfunction equation is
\begin{equation}
\label{eigenfunction_generalized_taboo}
	 - \frac{\sigma^2}{2} \frac{d^2 \varphi_n(x)}{d x^2} + \alpha \frac{\pi \sigma^2}{2L} \tan \left( \frac{\pi x}{2L} \right)   \frac{d \varphi_n(x)}{d x}  = \lambda_n(\alpha)  \varphi_n(x)
\end{equation}
where $\lambda_n(\alpha)$ is the n-th eigenvalue. Note that the lowest eigenvalue is $\lambda_0(\alpha) = 0$ (with $\varphi_0$  a constant function) so that the spectral gap, $\lambda_1(\alpha) - \lambda_0(\alpha)$, is equal to $\lambda_1(\alpha)$. The solution of this eigenvalues problem can be obtained by resorting to the so-called ultraspherical polynomials $G_n^{k}(x)$ (also known as Gegenbauer polynomials). Ultraspherical polynomials satisfy the differential equation~\cite{ref_Chapon}, 
\begin{equation}
\label{ultraspherical_polynomial_differential_equation}
	(1-x^2) f''(x) - (2 k +1 )x f'(x) + n(n+2 k) f(x) = 0 \, ,
\end{equation}
where $n$ is an integer and $k \ge -1/2$. A change of variable $x \mapsto \frac{2L}{\pi}\arcsin(x)$ shows that $G_n^{k}(\sin(\frac{\pi x}{2L}))$ is solution of the differential equation,
\begin{equation}
	  g''(x) - k \frac{\pi}{L} \tan \left( \frac{\pi x}{2L} \right) g'(x)  = - \left(\frac{\pi}{2L}\right)^2 n(n+2 k) g(x) \, .
\end{equation}
Consequently, by identifying the parameter $k$ with $\alpha$, the Gegenbauer polynomials are the solutions of the eigenvalue equation Eq.(\ref{eigenfunction_generalized_taboo}) with eigenvalues,
\begin{equation}
\label{eigenvalues_generalized_taboo}
 	\lambda_n(\alpha) = \frac{\sigma^2}{2}\left(\frac{\pi}{2L}\right)^2 n \left(n + 2 \alpha \right) \, ,
\end{equation}
and the spectral gap is,
\begin{equation}
\label{spectral_gap_generalized_taboo}
 	\lambda_1(\alpha) = \frac{\sigma^2}{2}\left(\frac{\pi}{2L}\right)^2 \left(1 + 2 \alpha \right) \, .
\end{equation}
\noindent Observe that when $\alpha = 1$, corresponding to the standard taboo case, we recover that the spectral gap is that given by~Eq.(\ref{spectral_gap_1D})
\begin{equation}
 	\lambda_1(\mathrm{taboo})= \frac{3 \sigma^2}{2}\left(\frac{\pi}{2L}\right)^2 .
\end{equation}
When $\alpha = 1 /2 $, corresponding to the sweetest taboo case, the spectral gap becomes
\begin{equation}
 	\lambda_1(\mathrm{sweetest ~taboo}) = \sigma^2 \left(\frac{\pi}{2L}\right)^2 ,
\end{equation}
and remains twice larger than the spectral gap of the reflected Brownian motion (Eq.(\ref{thm_reflected}) of appendix~\ref{appendix_4}). When $\alpha$ increases, which physically corresponds to more and more repulsive walls, the convergence towards the stationary density is faster, as indicated by~Eq.(\ref{spectral_gap_generalized_taboo}) and in accordance with Eq.(\ref{thm_comparison_spectral_gap}) of appendix~\ref{appendix_4}.

%%%%%%%%%%%%%%%%%%%%%%%%%%%%%%%%%%%%%%%%%%%%%%%%%%%%%%%%%%%%%%%%%%%%%%%%%%%%%%%%%%%%%%%%%%%%%%%%%%%%%%%%%%%%%%%%
%%																						%%
%%                 ASYMMETRIC DIFFUSION PROCESS IN AN INTERVAL                                                %%
%%																						%%
%%%%%%%%%%%%%%%%%%%%%%%%%%%%%%%%%%%%%%%%%%%%%%%%%%%%%%%%%%%%%%%%%%%%%%%%%%%%%%%%%%%%%%%%%%%%%%%%%%%%%%%%%%%%%%%%
\section{Asymmetric diffusion process in an interval }
\label{sec_asymmetric_diffusion}

%%%%%%%%%%%%%%%%%%%%%%%%%%%%%%%%%%%%%%%%%%%%%%%%%%%%%%%%%%%%%%%%%%%%%
%%                INVARIANT DENSITY                                %%
%%%%%%%%%%%%%%%%%%%%%%%%%%%%%%%%%%%%%%%%%%%%%%%%%%%%%%%%%%%%%%%%%%%%%
\subsection{Invariant density}
\label{subsec_invariant_density_with_drift}

The generalized taboo process, despite its advantages (closed-form expression of the stationary density, simulation easiness) suffers from an inherent drawback: It does not depend of the original drift of the underlying diffusion process. At a first sight this simplicity may be an advantage, but it does not reflect the expected behavior of a population described by an Ito diffusion with constant drift $\mu$ and constrained to stay in an interval. Indeed, for such a process, assuming that the walls have identical properties, it is expected that the invariant density get closer to the border in the drift direction 
%%(the left boundary for $\mu > 0$ and the right one for $\mu <0$)
. In some way, a conditioning \`{a} la Doob with respect to an event of zero probability, is so strong that it erases certain peculiarity of the initial process.\\

\noindent To overcome this difficulty, again we consider a (free) Ito diffusion with constant parameters $\mu$ and $\sigma$, driven by the stochastic differential equation,
\begin{equation}
	d X_t  = \mu \, dt +   \sigma dW_t \, .
\end{equation}
\noindent From this diffusion, we look for a process constrained to stay in a prescribed interval, says $[-L,L]$, of the form,
\begin{equation}
	d Z_t  = \left[\mu \, + \sigma^2 b(Z_t) \right] dt +   \sigma dW_t \, .
\end{equation}
\noindent The $\mu$ term ensures that the constrained process keeps its original drift characteristic while the $b(x)$ term, that needs to be specified, forces the process to remain within its boundaries. From the previous paragraph, we know that a necessary condition so that the process never leaves the interval is that the scale measure diverges at both ends, i.e. $S(-L,x] = S(L,x] = \infty $. For the left boundary, this quantity is given by~Eq.(\ref{speed_measure}),
\begin{equation}
 	S(-L,x]   =  \int_{-L}^x  e^{-  \frac{2 \mu}{\sigma^2} \eta }  e^{- 2 \int_0^{\eta} b(\xi)  d\xi }  d\eta \ge  e^{-  \frac{2 \mu}{\sigma^2} x } \int_{-L}^x e^{- 2 \int_0^{\eta} b(\xi)  d\xi }  d\eta \, .
\end{equation}
\noindent Assume that $b(x)$ behaves as $\alpha/(x+L)$ when $x \sim -L$, then the last integral diverges when $\alpha \ge 1/2 $. An identical reasoning applies to the right boundary. Since both barriers play a identical role, this suggests a $b(x)$ term that behaves as $\alpha/(x+L) + \alpha/(x-L)$ and thus a drift of the form
\begin{equation}
 	 \mu(x) = \mu + \alpha \sigma^2 \left( \frac{1}{x+L} + \frac{1}{x-L} \right) \, \mathrm{~with~~} \alpha \ge \frac{1}{2} \, .
\end{equation}
\noindent Before studying the general case, due to their importance, we examine two particular cases:

\begin{enumerate}
\item $\alpha = 1/2$ corresponding to the limit case (when $\alpha < 1/2$ the boundaries become permeable).
\item $\alpha = 1$ corresponding to the "standard" case.
\end{enumerate}

\noindent For both cases, we study the influence of the drift $\mu$ on the behavior of the conditioned process. We begin with the limit case. Recall the analytic expression of the scale function
\begin{equation}
      S(x) = \int_{0}^x s(\xi)  d\xi \, ; \qquad  s(\xi) = \exp \left\{- \int_{0}^{\xi} \frac{2 \mu(\eta)}{\sigma^2(\eta)}  d\eta \right\}  \, .
\end{equation}
\noindent Inserting the expression of the drift and diffusion coefficients in the preceding equation gives
\begin{equation}
\label{s_drift_interval}
 	s(\xi) =  \exp \left\{- \int_{0}^{\xi} \frac{2}{\sigma^2} \left[ \mu + \frac{\sigma^2}{2} \left( \frac{1}{x+L} + \frac{1}{x-L} \right) \right] d\eta \right\}  = L^2  \, \frac{ e^{- \frac{2 \mu \xi}{\sigma^2}} }{L^2-\xi^2}  \, .
\end{equation}
Next, to prove that $-L$ is an entrance boundary, we must verify that $S(-L,x]=\infty$ while $N(-L) < \infty$. The first part comes easily,
\begin{equation}
    S(-L,x] =  \int_{-L}^x s(\eta)  d\eta = L^2  \int_{-L}^x  \frac{ e^{- \frac{2 \mu \eta}{\sigma^2}} }{L^2-\eta^2} d\eta \, . \\    
\end{equation}
\noindent Since $ \frac{ e^{-\frac{2 \mu }{\sigma^2} x } }{L^2-x^2} \underset{x \to -L}{\sim\,}  \frac{L e^{\frac{2 \mu L}{\sigma^2} }}{2(x+L)}$ the preceding integral clearly diverges. It remains to verify that $N(-L)$ is finite. Recall that for any fixed $x \in [-L,L]$,

\begin{equation}
      \begin{aligned}
    N(-L) =  \int_{-L}^x  \left( \int_{-L}^{\xi} \frac{d\eta}{\sigma^2 s(\eta)} \right) s(\xi) d\xi 
	& = \frac{1}{\sigma^2} \int_{-L}^x \left( \int_{-L}^{\xi}  e^{\frac{2 \mu }{\sigma^2} \eta } (L^2-\eta^2) d\eta \right) \frac{ e^{-\frac{2 \mu }{\sigma^2} \xi } }{(L^2-\xi^2)}   d\xi \\ 
	& \le \frac{e^{\frac{4 \mu }{\sigma^2} L }}{\sigma^2} \int_{-L}^x \Big( \underbrace{\int_{-L}^{\xi}  (L^2-\eta^2)d\eta }_{\le L^2(\xi+L)}        \Big)   \frac{ d\xi }{(L^2-\xi^2)} \\
	& \le \frac{e^{\frac{4 \mu }{\sigma^2} L }}{\sigma^2} L^2 \int_{-L}^x  \frac{d\xi}{(L-\xi)}   \\
	& \le \frac{e^{\frac{4 \mu }{\sigma^2} L }}{\sigma^2} L^2 \log\left(\frac{2L}{L-x}\right) < \infty \, .
      \end{aligned}
\end{equation}

\noindent An identical reasoning can be made on the right boundary, and we can safely conclude that both boundaries are entrance.  Consequently the diffusion process starting inside $[-L,L]$ and driven by the stochastic differential equation
\begin{equation}
\label{SDE_new_process}
	d Z_t  = \left[\mu \, + \frac{\sigma^2}{2} \left( \frac{1}{Z_t-L} + \frac{1}{Z_t+L} \right) \right] dt +   \sigma dW_t \, ,
\end{equation} 
\noindent will stay forever in the interval $[-L,L]$. In addition, Eq.(\ref{stationary_density_simplified}) of appendix~\ref{appendix_3} provides the expression of the stationary density $\Psi(x)$. Inserting the expression of $s(x)$ (Eq.(\ref{s_drift_interval})) into  Eq.(\ref{stationary_density_simplified}) gives
\begin{equation}
\label{stationary_density_sweetest_drift_interval}
      \Psi(x)  = \frac{1}{ s(x) \int_{-L}^L \frac{d\eta}{ s(\eta)} } =
\frac{2 \mu^3 (L^2-x^2)e^{\frac{2 \mu x}{\sigma^2}}  }{\sigma^4 \left(2 \mu L \cosh\left(\frac{2 L \mu}{\sigma^2}\right) -\sigma^2 \sinh\left(\frac{2 L \mu}{\sigma^2}\right) \right)} \, .
\end{equation}
\noindent Note that in the absence of the original drift ($\mu = 0$), the stationary density reduces to,
\begin{equation}
\label{stationary_density_sweetest_drift_nul}
      \Psi(x)  = \frac{3}{4 L^3} (L^2-x^2) \, ,
\end{equation}
\noindent where as expected, the invariant density is a symmetric function with respect to $0$ (the middle of the interval).\\
%%since in this case boundaries play a symmetrical role.\\

Next, we examine the second case, when $\alpha = 1$, corresponding to boundaries that behave as "regular" walls. With this set of parameters the function $s(x)$ writes,
\begin{equation}
 	s(\xi) =  \exp \left\{- \int_{0}^{\xi} \frac{2}{\sigma^2} \left[ \mu + \sigma^2 \left( \frac{1}{\eta-L} + \frac{1}{\eta+L} \right) \right] d\eta \right\} 
            = L^4  \, \frac{ e^{-\frac{2 \mu \xi}{\sigma^2}} }{(L^2-\xi^2)^2} \, .
\end{equation}
\noindent From this expression, it is easy to check that the scale function diverges and that the double integral giving $N(-L)$ is finite. Like in the previous case, both boundaries are entrance. The stationary density is,
\begin{equation}
\label{stationary_density_sweetest_drift_alpha2_interval}
      \Psi(x)  = \frac{1}{ s(x) \int_{-L}^L \frac{d\eta}{ s(\eta)} }
 = \frac{2 \mu^5 (L^2-x^2)^2 e^{\frac{2 \mu x}{\sigma^2} }  }{-6 L \mu \sigma^8 \cosh\left(\frac{2 L \mu}{\sigma^2}\right) +  \sigma^6 (4 L^2 \mu^2 + 3 \sigma^4) \sinh\left(\frac{2 L \mu}{\sigma^2}\right)} \, .
\end{equation}
\noindent Again, in the absence of drift, the invariant density is a symmetric function with respect to the origin
\begin{equation}
\label{stationary_density_sweetest_drift_nul_alpha2}
      \Psi(x)  = \frac{15}{16 L^5}(L^2-x^2)^2 \, .\\
\end{equation}

\begin{figure}[h]
\centering
\includegraphics[width=3.5in,height=5.in,angle=-90]{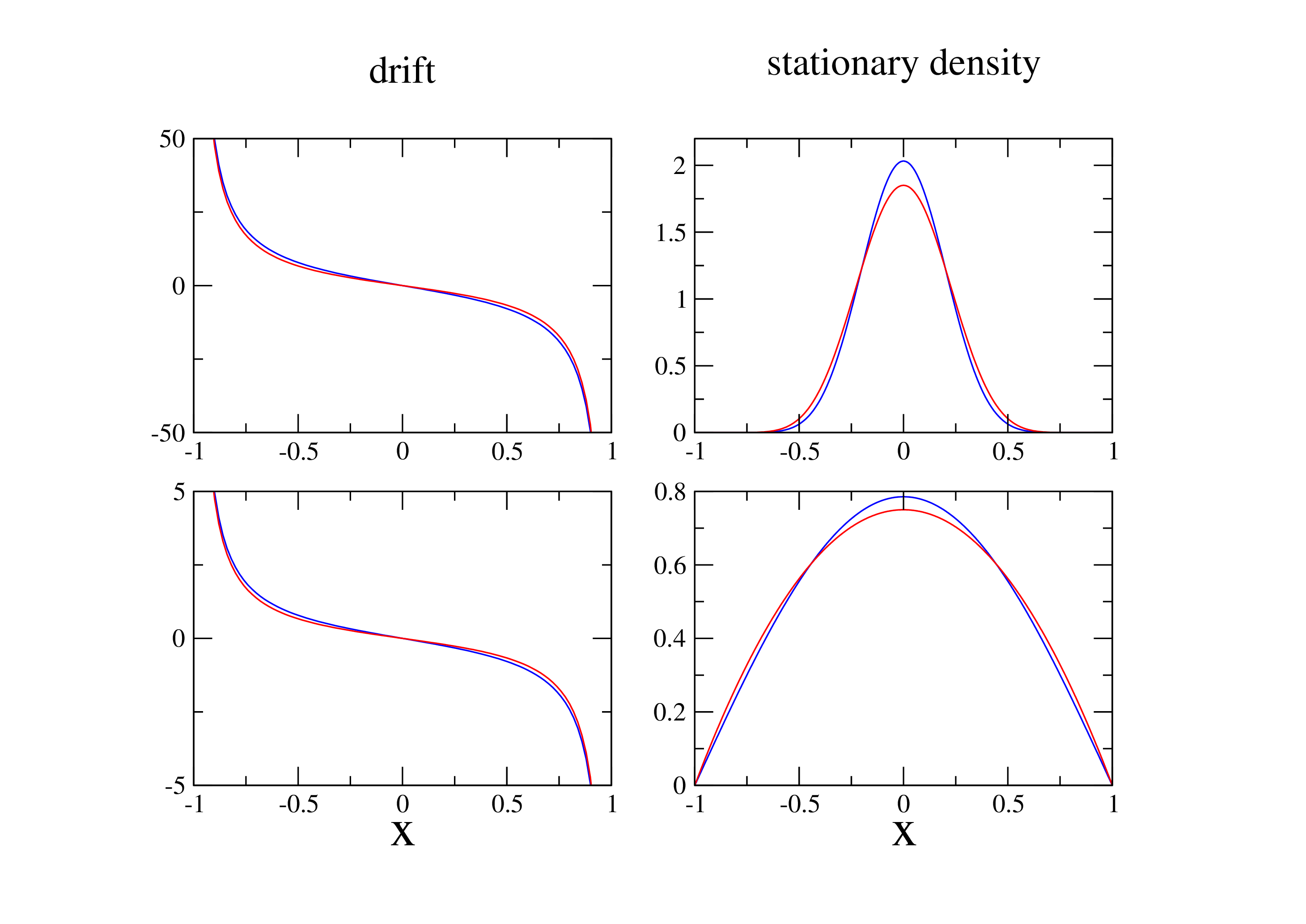}
\setlength{\abovecaptionskip}{15pt} 
\caption{Drifts of the generalized taboo process and the asymmetric diffusion as well as their corresponding invariant densities. Drift terms are plotted on the left column: blue lines correspond to generalized taboo processes and red lines to processes given by Eq.(\ref{SDE_generalized_new_process}) with $\mu=0$ and $\alpha=\beta$. The right column displays their corresponding invariant densities. The lower row corresponds to the sweetest taboo case ($\alpha=1/2$), the upper row corresponds to a drift 10 times larger.}
\label{fig3}
\end{figure}

It is not difficult to generalize the previous calculations for any value of $\alpha$, (with $\alpha \ge 1/2$ to assure entrance boundary conditions). Proceeding in the same way, we find that,
\begin{equation}
 	s(\xi)  = L^{4 \alpha} \, \frac{ e^{-\frac{2 \mu \xi}{\sigma^2}} }{(L^2-\xi^2)^{2 \alpha}}\, ,
\end{equation}
\noindent and the stationary density
\begin{equation}
\label{stationary_density_sweetest_drift_alphan_interval}
      \Psi(x)  = (L^2-x^2)^{2 \alpha} e^{\frac{2 \mu x}{\sigma^2} }  \frac{\Gamma \left( \frac{3}{2} + 2 \alpha \right)}{L^{4 \alpha +1} \sqrt{\pi} \Gamma \left(1 + 2 \alpha \right) _{0}\!F_{1}(\frac{3}{2} + 2 \alpha,\frac{L^2 \mu^2}{\sigma^4}) } \, ,
\end{equation}
\noindent where $_{0}\!F_{1}$ is the confluent hypergeometric function~\cite{ref_book_Abramowitz}. For $\mu=0$, the invariant density simplifies to the symmetrical expression
\begin{equation}
\label{stationary_density_sweetest_drift_nul_alphan}
      \Psi(x)  = (L^2-x^2)^{2 \alpha}  \frac{\Gamma \left( \frac{3}{2} + 2 \alpha \right)}{L^{4 \alpha +1} \sqrt{\pi} \Gamma \left(1 + 2 \alpha \right)  } \, .
\end{equation}
\noindent Figure~\ref{fig3} shows this invariant density as well as that of the asymmetric diffusion process (introduced just below).

\begin{figure}[h]
\centering
\includegraphics[width=3.in,height=4.in,angle=-90]{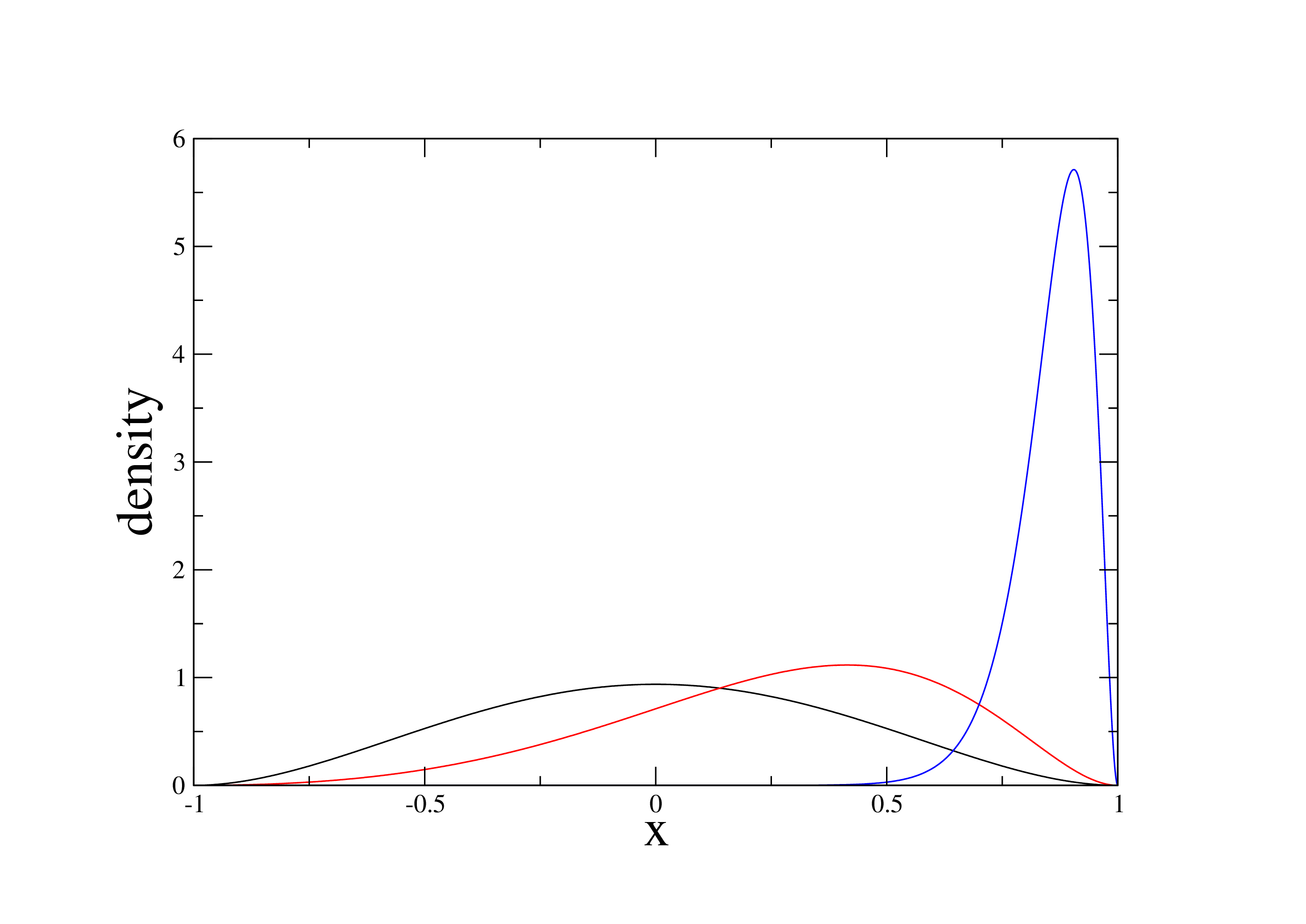}
\setlength{\abovecaptionskip}{15pt} 
\caption{Invariant density of the process described by the stochastic differential equation, $ d Z_t  = \left[\mu \, + \frac{1}{Z_t + L} + \frac{1}{Z_t - L} \right] dt +  dW_t $ in the interval $[-1,1]$ i.e. Eq.(\ref{SDE_generalized_new_process}) with the parameters $\alpha = \beta = \sigma = 1$. The invariant density is given by Eq.(\ref{stationary_density_sweetest_drift_asymmetric_interval}). The black line which is symmetric with respect to the origin corresponds to the driftless case $\mu=0$, red line: $\mu = 1 $, blue line: $\mu = 10 $. When the drift parameter increases, the invariant density is pushed towards the right wall significantly.}
\label{fig4}
\end{figure}

We can push a step further the generalization by allowing the walls to have different behaviors. The simplest form of the drift is the following:
\begin{equation}
	\mu(x) = \mu + \sigma^2 \left( \frac{\alpha}{x+L} + \frac{\beta}{x-L} \right) \, ,
\end{equation}
\noindent where as usual, $\alpha \ge 1/2$ and $\beta \ge 1/2$ to preserve entrance boundary conditions. The stochastic differential equation satisfies by the asymmetric diffusion process in an interval with different (entrance) boundaries is thus,
\begin{equation}
\label{SDE_generalized_new_process}
	d Z_t  = \left[\mu \, + \sigma^2 \left( \frac{\alpha}{Z_t + L} + \frac{\beta}{Z_t - L} \right) \right] dt +   \sigma dW_t \, .
\end{equation}
\noindent In the same vein, we can derive the stationary density of the process, which is
\begin{equation}
\label{stationary_density_sweetest_drift_asymmetric_interval}
      \Psi(x)  = \frac{(L+x)^{2 \alpha} (L-x)^{2 \beta} e^{\frac{2 \mu (L+x)}{\sigma^2} } \Gamma \left(2(1+\alpha+\beta) \right)} {(2L)^{1+2\alpha+2\beta} \Gamma \left(1+2 \alpha \right) \Gamma \left(1+2 \beta \right) ~_{1}\!F_{1}(1+2\alpha,2(1+\alpha+\beta),\frac{4 L \mu}{\sigma^2}) }  \, ,
\end{equation}
\noindent where $_{1}\!F_{1}$ is the Kummer confluent hypergeometric function~\cite{ref_book_Abramowitz}. Figure~\ref{fig4} shows this invariant density for various values of the drift $\mu$.

\noindent When $\mu = 0$, the stationary density reduces to
\begin{equation}
      \Psi(x)  = (L+x)^{2 \alpha} (L-x)^{2 \beta}  \frac{\Gamma \left(2(1+ \alpha + \beta) \right)}{(2L)^{1+2\alpha+2\beta} \Gamma \left(1+2 \alpha \right) \Gamma \left(1+2 \beta \right) } \, ,
\end{equation}
which is symmetric with respect to the origin only when $\alpha=\beta$, as expected. 

%%%%%%%%%%%%%%%%%%%%%%%%%%%%%%%%%%%%%%%%%%%%%%%%%%%%%%%%%%%%%%%%%%%%%
%%                  RATE OF CONVERGENCE                            %%
%%%%%%%%%%%%%%%%%%%%%%%%%%%%%%%%%%%%%%%%%%%%%%%%%%%%%%%%%%%%%%%%%%%%%
\subsection{Rate of convergence}
As we mentioned earlier, the rate of convergence towards the stationary density is a key quantity to describe an ergodic diffusion process in an interval. For sake of simplicity, we will focus on identical wall by considering a diffusion process in $[-L,L]$ characterized by its infinitesimal operator
\begin{equation}
\label{infinitesimal_generator_driftless_interval}
    \mathcal{L}.  =   \frac{\sigma^2}{2} \frac{d^2 .}{d x^2}  + \kappa \sigma^2 \left( \frac{1}{x+L} + \frac{1}{x-L}  \right)   \frac{d .}{d x} \, 
\end{equation}
where $\kappa \ge 1/2$ to insure that the process remains in $[-L,L]$ forever (assuming, of course, that the process starts inside  the interval). Again, the goal is to determine the spectral gap, i.e. the difference between the two smallest eigenvalues of the eigenvalue equation  $-\mathcal{L}\varphi_n(x) = \lambda_n(\kappa) \varphi_n(x)$. However, despite its apparent simplicity, we were not able to determine the analytical expression of the eigenvalues. 
In addition, it is well-known that a one-variable Fokker-Planck equation can always be transformed to a Schr\"odinger equation that has the same spectral gap~\cite{ref_book_Risken,ref_Pinsky_gap}. However, for the present drift, this strategy does not give a better result. Indeed, the transformation of the Fokker-Planck equation into a Schr\"odinger equation, detailed in appendix~\ref{appendix_5}, leads to,
\begin{equation}
	\mathcal{H}.  =   \frac{\sigma^2}{2}  \frac{d^2 .}{d x^2}  - V(x).  \, , 
\end{equation}
\noindent with a potential of the form
\begin{equation}
	 V(x) =  -\kappa \sigma^2 \frac{ \left( L^2 + x^2 - 2 \kappa x^2 \right)}{\left(L^2-x^2\right)^2} \,  .
\end{equation}
Unfortunately, this potential does not belong to any known family of exactly solvable potentials~\cite{ref_Cooper} even in the simplest case when $\kappa=1$. If the exact expression of the spectral gap seems out of reach, bounds are accessible thanks to some functional results obtained by Pinsky in 2005~\cite{ref_Pinsky_gap}. Pinsky's results are summarized (and adapted to the present study) in appendix~\ref{appendix_4}. Thanks to his results, spectral gap bounds will come quickly. Indeed, the drift 
\begin{equation}
     \mu(\kappa) = \kappa \sigma^2 \left( \frac{1}{x+L} + \frac{1}{x-L}  \right) \, ,
\end{equation}
\noindent is anti-symmetric and as such satisfies the conditions of Eq.(\ref{thm_reflected}) of appendix~\ref{appendix_4}. Hence, an immediate lower bound for the spectral gap is,
\begin{equation}
	\lambda_1(\mu) \ge \frac{\sigma^2}{2}\left(\frac{\pi}{2L}\right)^2 \, .
\end{equation}
Thanks to the results concerning the spectral gap of the generalized taboo process obtained  in the previous section~\ref{subsec_rate_of_convergence}, this lower bound can be significantly improved for certain values of $\kappa$. More precisely, let 

\begin{equation}
	\mu_T(\alpha) = -\alpha \frac{\pi \sigma^2}{2L} \tan \left( \frac{\pi x}{2L}  \right)  
\end{equation}
\noindent denotes the drift term of the generalized taboo process, for which the spectral gap is known (Eq.(\ref{spectral_gap_generalized_taboo})). The Eq.(\ref{thm_comparison}) of appendix~\ref{appendix_4} states that if $\mu(x) \ge \mu_T(x)$ on $[-L,0]$ and $\mu(x) \le \mu_T(x)$ on $[0,L]$ then: $\lambda_1(\mu) >  \lambda_1(\mu_T)$. Given the expressions of the different drifts, it means seeking $\kappa$  such that:

\begin{equation}
	\kappa \ge \alpha \frac{\pi}{4} \frac{\left(1-x^2\right)}{x} \tan\left(\frac{\pi x}{2} \right)  \mathrm{~for~} x\in[-1,1] .
\end{equation}
The function $f(x) = \frac{\pi}{4} \frac{\left(1-x^2\right)}{x} \tan\left(\frac{\pi x}{2} \right)$ is symmetric, and on the interval $[-1,1]$ has its maximum at the origin, which is $f(0) =\frac{\pi^2}{8}$. Consequently for $\alpha \ge 1/2$ and $\kappa \ge \alpha \frac{\pi^2}{8}$ the spectral gap of the process is greater than that of the generalized taboo process, which is $\lambda_1(\kappa) \ge \frac{\sigma^2}{2}\left(\frac{\pi}{2 L}\right)^2 \left(1 + 2 \alpha\right)$. In other words, we have the following result. \\

\noindent Consider a one-dimensional diffusion process in the interval $[-L,L]$ characterized by a variance coefficient $\sigma^2$ and a drift term of the form $\kappa \, \sigma^2 \left( \frac{1}{x+L} + \frac{1}{x-L}  \right)$.  If $\kappa \ge \frac{\pi^2}{16}$, then the spectral gap,

\begin{equation}
	\lambda_1(\kappa)  \ge \frac{\sigma^2}{8 L^2} \left(\pi^2 + 16 \, \kappa \right)\, .
\end{equation}

\noindent In particular the process converges faster than the taboo process towards its equilibrium density for a value of $\kappa \ge \frac{\pi^2}{8} \simeq 1.234$.

%%rev  \begin{thm}
%%rev  \label{thm_spectral_gap_diffusion_interval}
%%rev  Consider a one-dimensional diffusion process in the interval $[-L,L]$ characterized by a variance coefficient $\sigma^2$ and a drift term of the form $\kappa \, \sigma^2 \left( \frac{1}{x+L} + \frac{1}{x-L}  \right)$.  If $\kappa \ge \frac{\pi^2}{16}$, then the spectral gap:
%%rev  $$
%%rev     	\lambda_1(\kappa)  \ge \frac{\sigma^2}{8 L^2} \left(\pi^2 + 16 \, \kappa \right)\, .
%%rev  $$
%%rev  \noindent In particular the process converges faster than the taboo process towards its equilibrium density for a value of $\kappa \ge \frac{\pi^2}{8} \simeq 1.234$.
%%rev  \end{thm}

%%%%%%%%%%%%%%%%%%%%%%%%%%%%%%%%%%%%%%%%%%%%%%%%%%%%%%%%%%%%%%%%%%%%%
%%         MEAN EXIT TIME FROM AN INTERVAL                         %%
%%%%%%%%%%%%%%%%%%%%%%%%%%%%%%%%%%%%%%%%%%%%%%%%%%%%%%%%%%%%%%%%%%%%%
\subsection{Mean exit time from an interval}
\label{subsec_mean_exit_time}
\begin{figure}[h]
\centering
\includegraphics[width=3.5in,height=5.in,angle=-90]{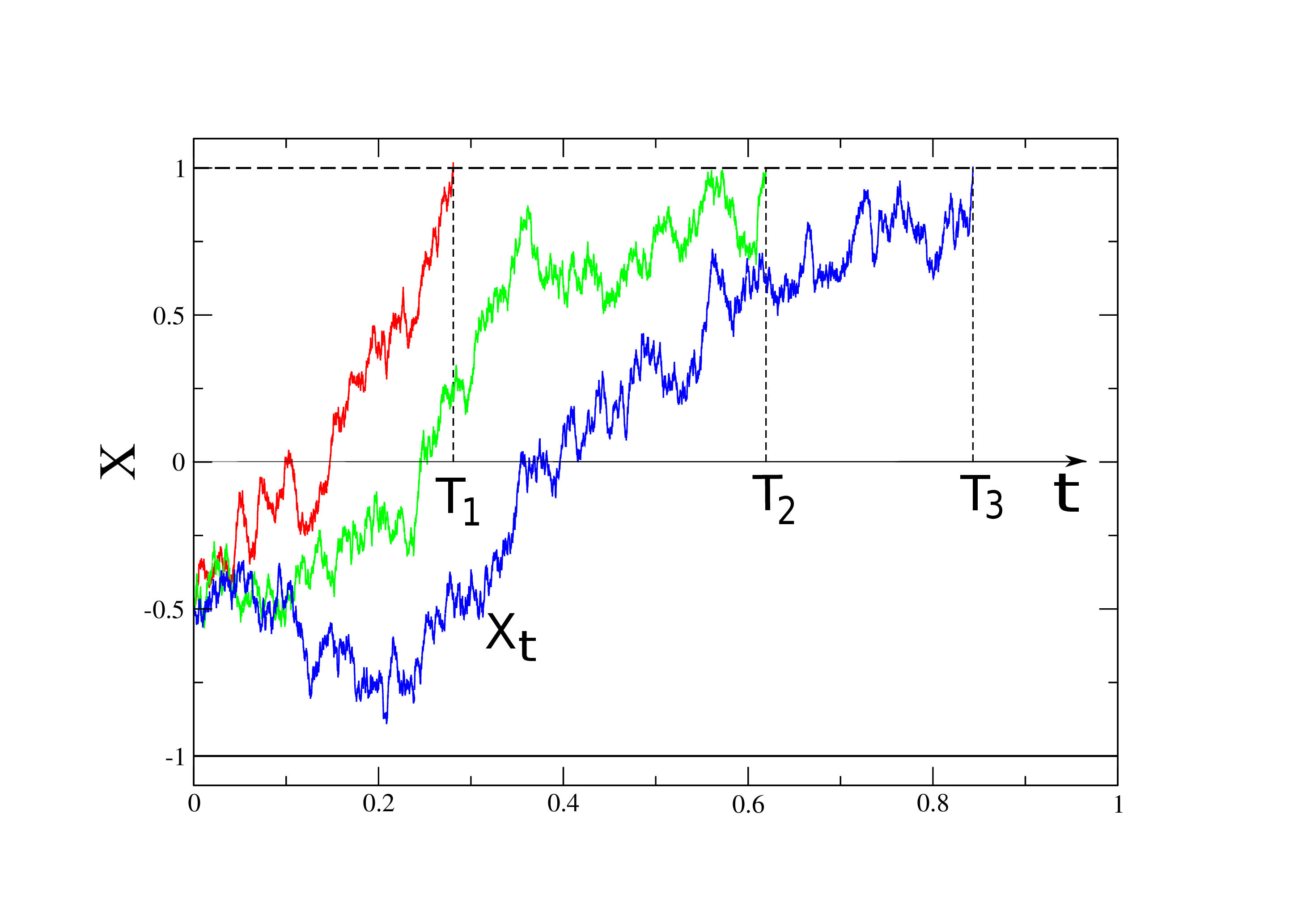}
\setlength{\abovecaptionskip}{15pt} 
\caption{Three realizations of the taboo process (without drift) in the interval $[-1,1]$ starting at $x_0 = -0.5$ and killed at $L=1$ as well as their corresponding first passage times $T_1$, $T_2$ and $T_3$.}
\label{fig5}
\end{figure}

To end our study of taboo processes and asymmetric diffusion processes in one dimension, we examine the mean exit time of these processes from an interval $[a,b]$ when one barrier becomes absorbing, the other one remaining an unreachable taboo state. To fix the ideas, let us consider that  $b$ is the absorbing boundary, and $a$ the taboo one. The process $X_t$ is killed (or stopped) when it reaches $b$ for the first time. Let $T$ be this first exit time. Some realizations of the process are shown in Fig.~\ref{fig5}.
We denote by $t(x)$ the mean exit time when the process starts from a point $x$ inside the domain, that is the functional 

\begin{equation}
\label{mean_exit_time_def}
      t(x) = \E [T | X_0 = x ] \, , \qquad a < x < b  .
\end{equation}

\noindent As in Karlin and Taylor's book~\cite{ref_book_Karlin}, without additional cost, we examine the expectation $w(x)$ of the more general functional $\int_0^T g(X_t) dt$, which is

\begin{equation}
\label{general_functional_def}
      w(x) = \E \left[\int_0^T g(X_t) dt | X_0 = x \right]  , \qquad a < x < b \, ,
\end{equation}
\noindent where $g(x)$ is a bounded continuous function. The mean exit time corresponds to the function $g(x)=1$. It is well-known that $w(x)$ satisfies the second order differential equation~\cite{ref_book_Karlin}

\begin{equation}
\label{general_functional_eq}
      \frac{1}{2} \sigma^2(x) \frac{d^2 w(x)}{dx^2} + \mu(x) \frac{d w(x)}{dx} = -g(x) \, .
\end{equation}

\noindent This equation is easier to solve in its canonical form. Once again, we resort to the useful scale function $S(x)$ and the speed density $m(x)$, (both functions are defined through  Eqs.(\ref{scale_function_and_more}) in appendix~\ref{appendix_3}). Then Eq.(\ref{general_functional_eq}) becomes

\begin{equation}
\label{general_functional_eq_canonical}
      \frac{1}{2 m(x)} \frac{d }{dx} \left[ \frac{dw(x)}{dS(x)} \right]= -g(x) \, .
\end{equation}

\noindent Denoting by $C_1$ and $C_2$ the constants of integration, we get after two successive integrations,

\begin{equation}
\label{general_functional_sol_tempo}
      w(x) = -2  \int_a^x \left[ \int_a^{\eta} g(\xi) m(\xi) d\xi \right] dS(\eta) + C_1 \left[ S(x)-S(a) \right] + C_2 \, .
\end{equation}

\noindent The boundary conditions must now be specified. They are identical to those of the probability density $P(y,t| X_0 = x)$, of the stochastic process $X_t$ starting at $t=0$ with $X_0 = x$ and reaching  $y$ at time $t$~\cite{ref_book_Karlin}. Therefore, we have $w(b) = 0$ since $b$ is an absorbing boundary, and $w'(a) = 0$ since the net flow of the probability current is zero on an entrance boundary, no particle being able to cross it.
The boundary condition $w'(a)=0$ implies that $C_1=0$, and the boundary condition $w(b)=0$ requires (after inverting the order of integration) that $C_2 = 2 \int_a^b \left[ S(b)-S(\xi) \right] g(\xi) m(\xi) d\xi$. Finally, we obtain 

\begin{equation}
\label{general_functional_sol_final}
      w(x) = -2  \int_a^x \left[ S(b)-S(x) \right] g(\xi) m(\xi) d\xi + 2 \int_x^b \left[ S(b)-S(\xi) \right] g(\xi) m(\xi) d\xi \, , 
\end{equation}

\noindent and for the mean exit time, when $g(x)=1$, 

\begin{equation}
\label{mean_exit_time_sol_final}
      t(x) = -2  \int_a^x \left[ S(b)-S(x) \right] m(\xi) d\xi + 2 \int_x^b \left[ S(b)-S(\xi) \right]  m(\xi) d\xi \, .
\end{equation}

\noindent The general framework being ready, we return to the processes we wish to study. First, we begin with the taboo process in the semi-infinite state space $[a,+\infty)$, $a$ being, as we mentioned, the taboo state. With only one taboo state, the taboo process was studied by Knight~\cite{ref_Knight} (see also~\cite{ref_Pinsky}), and has parameters,

%%\begin{equation}
%%\label{infinitesimal_generator_driftless_semi_infinite}
%%    \mathcal{L}.  =   \frac{\sigma^2}{2} \frac{d^2 .}{d x^2}  +  \frac{\sigma^2}{x-a} \frac{d .}{d x} \, 
%%\end{equation}

\begin{equation}
\label{taboo_driftless_semi_infinite}
  \left\{
      \begin{aligned}
        \sigma(x) & = \sigma   \\
        \mu(x)    & = \frac{\sigma^2}{x-a} \, . \\
      \end{aligned}
    \right.
\end{equation}

\noindent Note that when $a=0$, the state space is the semi-infinite positive axis and the generator $(\sigma^2/2) (d^2 ./d x^2)  +  (\sigma^2/x) (d ./d x)$ is that of the well-known Bessel process, i.e. the radial part of a three-dimensional Brownian motion~\cite{ref_book_Karlin}. For the asymmetric process that we introduced in section~\ref{sec_asymmetric_diffusion}, with a single barrier, its parameters are (with $\alpha > 1/2$ to insure that $a$ is still an entrance boundary),

\begin{equation}
\label{process_semi_infinite}
  \left\{
      \begin{aligned}
        \sigma(x) & = \sigma   \\
        \mu(x)    & = \mu  + \alpha \frac{\sigma^2}{x-a} \, . \\
      \end{aligned}
    \right.
\end{equation}

\noindent Therefore, the taboo process with a single barrier is a special case of the asymmetric process, with $\mu = 0$ and $\alpha =1$. Since both processes can be treated in the same way, we collect the results with one calculus. We start with the driftless case, thus assuming $\mu =0$. With the parameters $\sigma(x)=\sigma$ and $\mu(x)=\alpha \sigma^2/(x-a)$ the scale function $S(x)$ and the speed density $m(x)$ 
%(given by Eqs.(\ref{scale_function_and_more}))
are

\begin{equation}
\label{taboo_driftless_semi_infinite_sacle_and_speed_fucntions}
	S(x) = \frac{(x-a)^{1-2 \alpha }}{1-2 \alpha } \qquad \mathrm{and} \qquad m(x) = \frac{(x-a)^{2 \alpha }}{\sigma ^2} \, .
\end{equation}

\noindent Reporting these expressions in Eq.(\ref{mean_exit_time_sol_final}), we obtain the mean exit time through $b$

\begin{equation}
\label{mean_exit_time_generalized_taboo_driftless}      
      t_b(x) = \frac{(b-x)(b+x-2a)}{(2 \alpha +1) \sigma ^2} .
\end{equation}

\noindent For the case of two symmetric barriers, $a=-L$ and $b=L$,

\begin{equation}
\label{mean_exit_time_generalized_taboo_driftless_sym}      
      t_L(x) = \frac{(L-x) (3 L+x)}{(2 \alpha +1) \sigma ^2} \, .
\end{equation}

\noindent From the preceding equation, we have immediately the following results:

\begin{equation}
\label{mean_exit_time_taboo_and_sweetest_sym} 
  \left\{
      \begin{aligned}
       &\text{taboo~} (\alpha=1):  &t_L(x) = \frac{(L-x) (3 L+x)}{3 \sigma ^2}
       \\
       &\text{sweetest taboo~} (\alpha=1/2):  &t_L(x) = \frac{(L-x) (3 L+x)}{2 \sigma ^2} \, .
      \end{aligned} 
    \right.
\end{equation}

\noindent These two mean exit times are to be compared to the mean exit time of the reflected Brownian motion (with one reflecting wall). The average exit time of the reflected Brownian motion is well-known~\cite{ref_book_Pavliotis,ref_book_Gillespie}

\begin{equation}
\label{mean_exit_time_brownian_reflecting}      
      t_L^{ref}(x) = \frac{(L-x) (3 L+x)}{\sigma ^2} \,  .
\end{equation}

\noindent Therefore, the average exit time of the reflected Brownian process is $(2 \alpha +1)$ times larger than that of the generalized taboo process. Since $\alpha>1/2$, we can conclude that the generalized taboo particle spends less time in the interval than the pure reflected Brownian motion. On physical grounds, we expected this behavior for the taboo process. Indeed, 
remember that the taboo process behaves like the reflected Brownian motion near the boundary. Furthermore, for the taboo process, a forward drift 
%(in $\sim 1/(x-L)$) 
exists everywhere in the domain. This drift pushes the process towards the absorbing boundary whereas the reflected Brownian motion evolves freely out of the reflecting wall. Thus, in average, the taboo process is expected to hit the absorbing boundary faster than the reflected Brownian motion. It is more surprising that this result remains true for the sweetest taboo process. \\

\noindent Now, we treat the general case, assuming that $\mu \ne 0$ and beginning with the asymmetric process with $\alpha=1$. For sake of simplicity, we consider the case of a symmetric interval $[-L,L]$. With the set of parameters, $\sigma(x)=\sigma$ and $\mu(x)=\mu + \sigma^2/(x-L)$, the scale function $S(x)$ and the speed density $m(x)$ are

\begin{equation}
\label{taboo_semi_infinite_sacle_and_speed_fucntions}
	S(x) =  - \frac{e^{-\frac{2 \mu  x}{\sigma ^2}}}{L+x} -\frac{2 \mu  e^{\frac{2 L \mu }{\sigma ^2}} \mathrm{Ei}\left(-\frac{2 (L+x) \mu }{\sigma ^2}\right)}{\sigma ^2} \qquad \mathrm{and} \qquad m(x) = \frac{(x+L)^2 e^{\frac{2 \mu  x}{\sigma ^2}}}{\sigma ^2} 
\end{equation}

\noindent where $\mathrm{Ei}(x)$ is the exponential integral function. Reporting these expressions in Eq.(\ref{mean_exit_time_sol_final}) and performing the integration gives the mean exit time through $L$,

\begin{equation}
\label{mean_exit_time_taboo}      
	\begin{aligned}
      t(x) & = \frac{1}{4 \mu ^3 L (L+x)} \left[ 4 \mu ^2 L (L^2-x^2) +\sigma ^4 x \left(e^{-\frac{4 \mu  L}{\sigma ^2}}-1\right) + \sigma ^4 L \left(e^{-\frac{4 \mu  L}{\sigma ^2}}-2 e^{-\frac{2 \mu  (L+x)}{\sigma ^2}}+1\right) \right]  \\
           & + \frac{\sigma ^2}{\mu^2} \left[ \text{Ei}\left(-\frac{4 L \mu }{\sigma ^2}\right)-\text{Ei}\left(-\frac{2 (L+x) \mu }{\sigma ^2}\right) +\log\left( \frac{1}{2}+\frac{x}{2L} \right) \right] \, .
	\end{aligned}
\end{equation}

\noindent Thanks to this approach, it is straightforward to obtain the mean exit time of the asymmetric process with $\alpha=1/2$. With the parameters, $\sigma(x)=\sigma$ and $\mu(x)=\mu + \sigma^2/(2(x+L))$ we get

\begin{equation}
\label{mean_exit_time_sweetest_taboo}      
      t(x) = \frac{L-x}{\mu } + \frac{\sigma ^2}{2 \mu ^2} \left[ \mathrm{Ei}\left(-\frac{4 L \mu }{\sigma ^2}\right)-\mathrm{Ei}\left(-\frac{2 (L+x) \mu }{\sigma ^2}\right) +\log\left( \frac{1}{2}+\frac{x}{2L} \right)\right] .
\end{equation}

\noindent Lastly, in the same vein, we give the mean exit time of the reflected Brownian motion with drift ($\sigma(x)=\sigma$ and $\mu(x)=\mu$), 
\begin{equation}
\label{mean_exit_time_brownian_drift}      
      t(x) = \frac{L-x}{\mu } + \frac{\sigma ^2}{2 \mu ^2} \left[ e^{-\frac{4 \mu  L}{\sigma ^2}}-e^{-\frac{2 \mu  (L+x)}{\sigma ^2}} \right] .
\end{equation}

\begin{figure}[h]
\centering
\includegraphics[width=3.in,height=6.in,angle=-90]{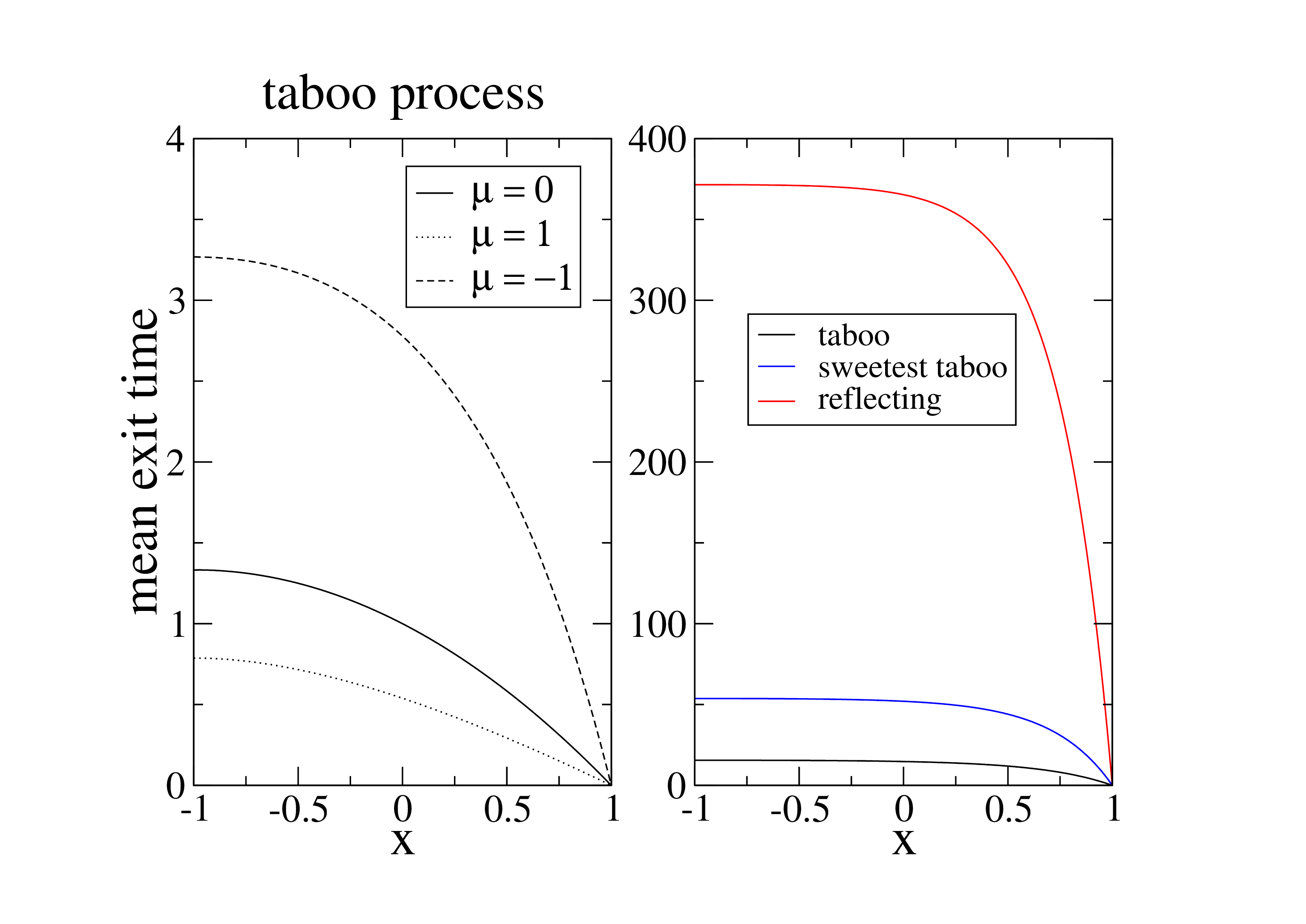}
\setlength{\abovecaptionskip}{15pt} 
\caption{Left: Mean exit time of the asymmetric process with $\alpha =1$ for three different values of the drift ($\mu=0, 1, -1$). 
Right: Mean exit time of different processes, all with the same negative drift $\mu = -2$. Black line: mean exit time of the asymmetric process with $\alpha =1$ given by Eq.(\ref{mean_exit_time_taboo}). Blue line: mean exit time of the asymmetric process with $\alpha =1/2$ given by Eq.(\ref{mean_exit_time_sweetest_taboo}).  Red line: mean exit time of the reflected diffusion  given by Eq.(\ref{mean_exit_time_brownian_drift}). All curves exhibit the same behavior: the mean exit time is almost constant except when the initial position approaches the absorbing boundary $L=1$. This plateau-shape curve is accentuated when $\mu$ becomes more negative.}
\label{fig6}
\end{figure}

\noindent The behavior of the different mean exit times are plotted on Fig.~\ref{fig6} for various values of the drift $\mu$. 
\noindent When $\mu$ becomes more negative, all curves tend towards a plateau-shape function that vanishes to zero when the initial position is close to the absorbing wall. Indeed, a very negative drift term will strongly push the process towards the reflecting boundary where it will remain for a very long time. Regarding the average exit time, it is as if the process started on the entrance boundary. Unless the process starts very near (or on) the absorbing boundary, in which case the exit probability becomes significant and the average exit time decreases.

%%%%%%%%%%%%%%%%%%%%%%%%%%%%%%%%%%%%%%%%%%%%%%%%%%%%%%%%%%%%%%%%%%%%%%%%%%%%%%%%%%%%%%%%%%%%%%%%%%%%%%%%%%%%%%%%
%%																						%%
%%                 TWO DIMENSIONAL TABOO PROCESSES                                                            %%
%%																						%%
%%%%%%%%%%%%%%%%%%%%%%%%%%%%%%%%%%%%%%%%%%%%%%%%%%%%%%%%%%%%%%%%%%%%%%%%%%%%%%%%%%%%%%%%%%%%%%%%%%%%%%%%%%%%%%%%
\section{Two dimensional taboo process}
\label{sec_Taboo2D}

\begin{figure}[h]
\centering
\includegraphics[width=3.5in,height=3.5in,angle=-90]{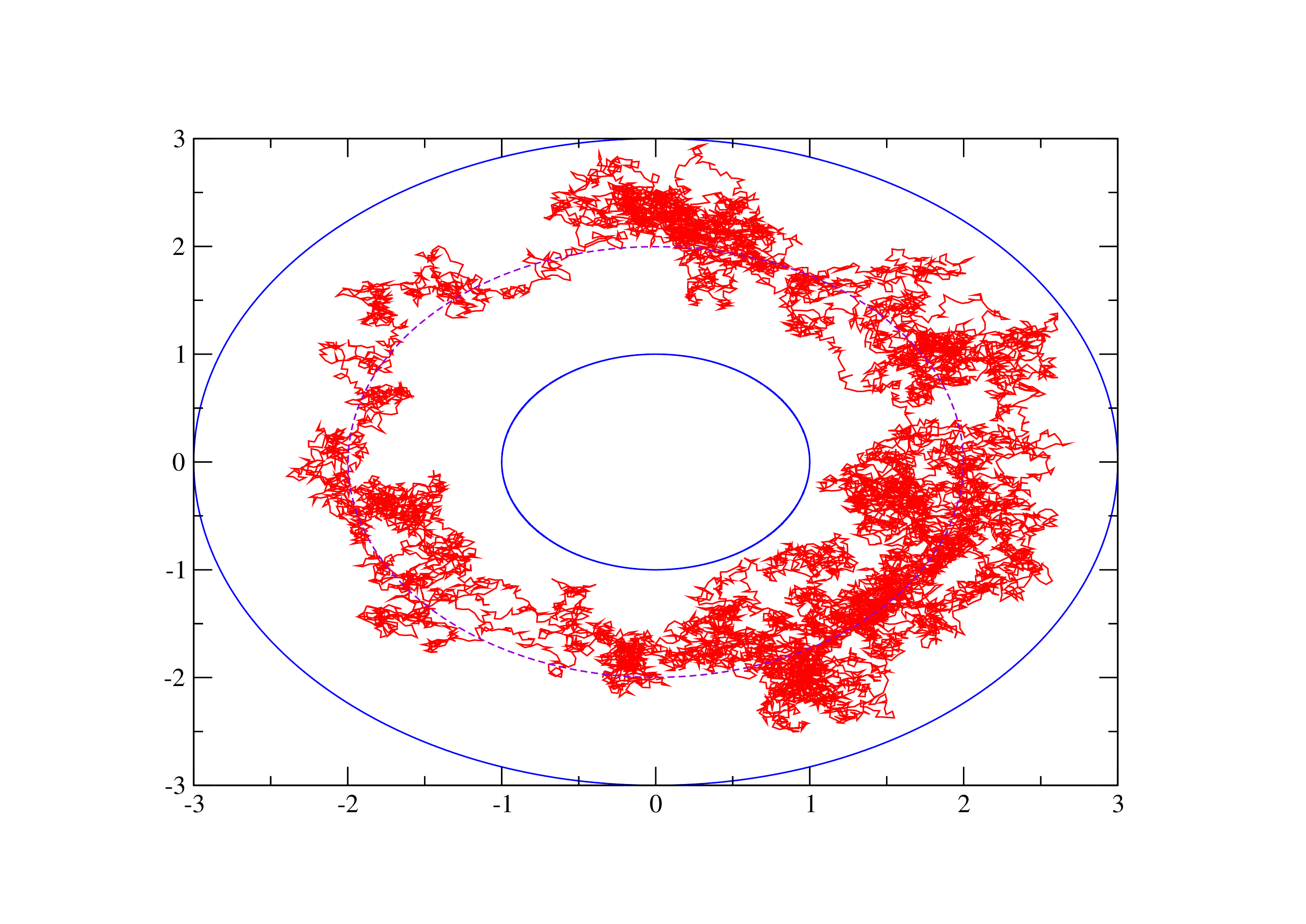}
\setlength{\abovecaptionskip}{15pt} 
\caption{Red line: a realization of the two-dimensional taboo process in a circular annulus. The purple line plots the mean radius of the paths $\E[r^*_{\infty}] \simeq 1.997$.}
\label{fig7}
\end{figure}

In this paragraph, we consider taboo processes in two-dimensional bounded domains. Being in two dimensions offers richer possibilities than the geometry of an interval, but leads also to technical difficulties. In the following, we consider a taboo process evolving inside a circular annulus, a non-convex region bounded by two concentric circles of radius $a$ and $b$,  $\mathcal{D}_A = {r \in R^2 : a < |r| < b}$. Figure~\ref{fig7} shows a typical path of the two-dimensional taboo process in this geometry. 
The annulus is of special interest since it allows to study the winding properties of the process. More generally, studying Brownian paths with topological constrains (here a hole) is an important subject in many fields of physics (physics of polymers~\cite{ref_Rudnick_Hu,ref_Houchmandzadeh}, fluxlines in superconductors~\cite{ref_Drossel}) and  mathematics~\cite{ref_Spitzer,ref_Pitman,ref_Belisle}. Besides, taboo processes in a circular annulus extended in some way the recent results exposed in the article of Kundu, Comtet and Majumdar concerning the properties 
%winding statistics 
of a Brownian particle on a ring~\cite{ref_Kundu}.\\

\noindent For sake of simplicity let us consider a pure two-dimensional Brownian motion. By taking, $\bm{\mu}(\bm{x}) = \left(\! \begin{array}{ll}  0 \\
      0 
\end{array} \! \right) $ and $\bm{\sigma}(\bm{x}) = \left(\! \begin{array}{ll} 1 & 0 \\
      0 & 1
\end{array} \! \right) $, the associated two-dimensional taboo process $ \bm{X}^*_t = \left(\! \begin{array}{ll} X_{1t}^*  \\
      X_{2t}^* 
\end{array} \! \right) $ in a circular annulus becomes (Eq.(\ref{taboo_SDE}))

\begin{equation}
\label{taboo_SDE_2D}
	\begin{pmatrix} 
    dX_{1t}^*  \\ 
    dX_{2t}^*
  \end{pmatrix} =  \begin{pmatrix}  
    \partial_{X_{1t}^*} \varphi_1(X_{1t}^*,X_{2t}^*) /  \varphi_1(X_{1t}^*,X_{2t}^*)  \\ 
    \partial_{X_{2t}^*} \varphi_1(X_{1t}^*,X_{2t}^*) /  \varphi_1(X_{1t}^*,X_{2t}^*) 
  \end{pmatrix}  dt + \begin{pmatrix} 
    dW_{1t}  \\ 
    dW_{2t}
  \end{pmatrix}    \, ,
\end{equation}

\noindent where $W_{1t}$  and $W_{2t}$ are two independent standard Brownian motions and where $\varphi_1(x_1,x_2)$ is the first eigenfunction of the operator $-\mathcal{L} = -1/2 \Delta_2$ (minus half of the 2-dimensional Laplacian) on the circular annulus domain $\mathcal{D}_A$ with Dirichlet boundary conditions on $\partial \mathcal{D}_A$. Due to the radial symmetry of the geometry, the lowest eigenfunction of the operator $-1/2 \Delta_2$ depends only on the radial component $r=\sqrt{x_1^2+x_2^2}$. Since the radial component of $-\frac{1}{2} \Delta_2$ is $- \frac{1}{2}\left(\frac{d^2}{dr^2} + \frac{1}{r} \frac{d}{dr}\right)$, the equation satisfies by the eigenfunction $\varphi_1(x_1,x_2) = \varphi_1(r)$ is 

\begin{equation}
     \frac{d^2 \varphi_1(r)}{dr^2} + \frac{1}{r} \frac{d \varphi_1(r)}{dr} = -2 \lambda \varphi_1(r).
\end{equation}

\noindent 
The solution of this equation is expressed in terms of the Bessel function of the first kind $J_0(r)$ and the Bessel function of the second kind $Y_0(r)$, and two constants $C_0$ and $C_1$,

\begin{equation}
\label{taboo_2D_phi_1_start}
      \varphi_1(r) = C_0 J_0(\sqrt{2 \lambda} r) +  C_1 Y_0(\sqrt{2 \lambda} r).
\end{equation}

\noindent Boundary conditions $\varphi_1(a) = \varphi_1(b) = 0$,

\begin{equation}
  \left\{
      \begin{aligned}
       C_0 J_0(a \sqrt{2 \lambda}) + C_1 Y_0(a \sqrt{2 \lambda}) &= 0
       \\
       C_0 J_0(b \sqrt{2 \lambda}) + C_1 Y_0(b \sqrt{2 \lambda}) &= 0 \\
      \end{aligned}
    \right.
\end{equation}

\noindent imply that,
\begin{equation}
      \begin{vmatrix}
J_0(a \sqrt{2 \lambda}) & Y_0(a \sqrt{2 \lambda}) \\
J_0(b \sqrt{2 \lambda}) & Y_0(b \sqrt{2 \lambda})
      \end{vmatrix} = 0   \Leftrightarrow    J_0(a \sqrt{2 \lambda})Y_0(b \sqrt{2 \lambda}) - J_0(b \sqrt{2 \lambda})Y_0(a \sqrt{2 \lambda}) = 0 \,  .
\end{equation}
\noindent For fixed numerical values of $a$ and $b$, the first roots of the preceding equation can be found in~\cite{ref_book_Carslaw} or is given by a mathematical software like Mathematica. The first root corresponds to the smallest eigenvalue and from now on $\lambda$ is fixed to this value. Up to a normalization constant $C_0$, $\varphi_1(r)$ writes,
\begin{equation}
\label{phi_1_polar_intermediate}
      \varphi_1(r) = C_0 \left[  J_0(\sqrt{2 \lambda} r) - \frac{J_0(\sqrt{2 \lambda} a)}{Y_0(\sqrt{2 \lambda} a)} Y_0(\sqrt{2 \lambda} r) \right],
\end{equation}
\noindent and,
\begin{equation}
      \frac{\bm{ \nabla} \varphi_1(r)}{\varphi_1(r)} = \underbrace{ \sqrt{2 \lambda} \left[ \frac{J_0(\sqrt{2 \lambda} a) Y_1(\sqrt{2 \lambda} r) - J_1(\sqrt{2 \lambda} r) Y_0(\sqrt{2 \lambda} a) }{J_0(\sqrt{2 \lambda} r)Y_0(\sqrt{2 \lambda} a) - J_0(\sqrt{2 \lambda} a)Y_0(\sqrt{2 \lambda} r) }  \right] }_{B(r)} \frac{\bm{r}}{r} \, .
\end{equation}
\noindent Converting to polar coordinates, $X_{1t}^* = r^*_t \cos(\theta^*_t)$ and $X_{2t}^* = r^*_t \sin(\theta^*_t)$ and introducing the scalar function $B(r)$, the stochastic differential equation Eq.(\ref{taboo_SDE_2D}) becomes,

\begin{equation} 
\label{taboo_SDE_2D_polar_start}
  \left\{
      \begin{aligned}
      d (r^*_t \cos(\theta^*_t))  & =  B(r^*_t) \cos(\theta^*_t) dt +  dW_{1t}  \\
      d (r^*_t \sin(\theta^*_t))  & =  B(r^*_t) \sin(\theta^*_t) dt +  dW_{2t} \, . \\
      \end{aligned}
    \right.
\end{equation}

\noindent Since $B$ is a function of $r$ only, we convert to polar coordinate the stochastic differential equation Eq.(\ref{taboo_SDE_2D_polar_start}). Applying multidimensional Ito's formula~\cite{ref_book_Schuss_2} to $r^*_t = \sqrt{(X_{1t}^*)^2 + (X_{2t}^*)^2}$ and $\theta^*_t = \arctan\left(X_{2t}^* / X_{1t}^* \right)$ we get that,

\begin{equation} 
\label{taboo_SDE_2D_polar_intermediate}
  \left\{
      \begin{aligned}
      d r^*_t      & = \left( B(r^*_t) + \frac{1}{2 r^*_t} \right) dt +\cos \theta^*_t \, dW_{1t} + \sin \theta^*_t \, dW_{2t} \\
      d \theta^*_t & = \frac{1}{r^*_t} \left(-\sin \theta^*_t \, dW_{1t} + \cos \theta^*_t \, dW_{2t} \right)  \, . \\
      \end{aligned}
    \right.
\end{equation}

\noindent Moreover, since the solutions $W^r_t$ and $W^{\theta}_t$ of the equations,
\begin{equation}
  \left\{
      \begin{aligned}
       d W^r_t         & =  \cos \theta^*_t \, dW_{1t} + \sin \theta^*_t \, dW_{2t} 
       \\
       d W^{\theta}_t  & =  -\sin \theta^*_t \, dW_{1t} + \cos \theta^*_t \, dW_{2t} \\
      \end{aligned} 
    \right.
\end{equation}

\noindent are two independent standard Brownian motions~\cite{ref_book_Schuss_2}, we can write the equations Eqs.(\ref{taboo_SDE_2D_polar_intermediate}) in the more compact form,
\begin{equation}
\label{taboo_SDE_2D_polar_final}
  \left\{
      \begin{aligned}
      d r^*_t      & = \left( B(r^*_t) + \frac{1}{2 r^*_t} \right) dt + d W^r_t \\
      d \theta^*_t & = \frac{ d W^{\theta}_t}{r^*_t}  \, , \\
      \end{aligned}  
    \right. \, 
\end{equation}

\noindent with the chosen initial conditions,
\begin{equation}
\label{taboo_SDE_2D_polar_final_initial_conditions}
  \left\{
      \begin{aligned}
      r^*_0 & = r_0 \qquad \mathrm{with~~} a < r_0 < b\\
      \theta^*_0 & = 0 \, . \\
      \end{aligned}  
    \right. .
\end{equation}

\noindent The strategy to solve this couple of stochastic differential equations consists in solving the equation on $r^*_t$ (which is independent of $\theta^*_t$), and then replacing this solution in the second equation related to $\theta^*_t$. Unfortunately, due to a drift term involving Bessel functions, a closed-form for $r^*_t$ is beyond reach. Nevertheless, it is still possible to study the behavior of the process in the long time limit. Indeed, since both boundaries are entrance, the process is ergodic and, as such, has converged (exponentially) to an invariant density for times greater than $\tau$, where $\tau$ is the inverse of the spectral gap. Once the process has converged, $r^*_t$ is distributed according to the density probability function $\Psi(r)$ which is time independent. We call $r^*_{\infty}$ this random variable. The solution of the stochastic differential equation~Eq.(\ref{taboo_SDE_2D_polar_final}) with the initial condition~Eq.(\ref{taboo_SDE_2D_polar_final_initial_conditions}) is
\begin{equation}
      \theta^*_t  =  \int_0^t \frac{ d W^{\theta}_u}{r^*_u}  \, . \\
\end{equation}
\noindent By construction $\theta^*_t$  is a martingale~\cite{ref_book_Karlin} and 
\begin{equation}
\label{taboo_SDE_2D_Ito_1}
	\E\left[\theta^*_t\right] = \E\left[\theta^*_0\right] = 0 \, .
\end{equation}
%% = \E\left[\int_0^t \frac{ d W^{\theta}_u}{r^*_u}  \right]
\noindent Moreover, Ito's isometry states that
\begin{equation}
      \E\left[\theta^{*2}_t \right] = \E\left[\left(\int_0^t \frac{ d W^{\theta}_u}{r^*_u} \right)^2 \right]  = \E\left[\int_0^t \left( \frac{1}{r^*_u} \right)^2 du \right] \, . \\
\end{equation}
\noindent Now, assume that the long time limit is reached, i.e. that $t \gg \tau$. Then, from the previous equation,
\begin{equation}
      \E\left[\theta^{*2}_t \right]  = \E\left[\int_0^{\tau} \left( \frac{1}{r^*_u} \right)^2 du \right] + \E\left[\int_{\tau}^t \left( \frac{1}{r^*_u} \right)^2 du \right] \, . \\
\end{equation}
\noindent The first integral is finite and bounded by $\tau/a^2$, and the integrand of the second integral is independent of $t$, thus
\begin{equation}
      \E\left[\theta^{*2}_t \right]  = \mathrm{finite~term} + \E \left[\left( \frac{1}{r^*_{\infty}} \right)^2 \right] (t - \tau) \, , \\
\end{equation}
\noindent or equivalently,
\begin{equation}
\label{taboo_SDE_2D_Ito_2}
      \E\left[\theta^{*2}_t \right] \underset{t \to \infty}{\sim}   \E \left[ \frac{1}{r^{*2}_{\infty}}  \right] t \, .
\end{equation}

From Eqs.(\ref{taboo_SDE_2D_Ito_1}) and (\ref{taboo_SDE_2D_Ito_2}) along with the initial condition $\theta^*_0 = 0 $, it follows from Levy's characterization of Brownian motion~\cite{ref_book_Revuz_Yor,ref_book_DelMoral} that in the limit of long times, $\theta^*_t$ is a Brownian motion whose variance is given by\footnote{Recall that the invariant density is given by~Eq.(\ref{Invariant_density_Pinsky}), $\Psi(r) =\varphi_1(r) \varphi_1^{\dagger}(r)$, and since the operator $-1/2 \Delta_2$ with Dirichlet conditions on $r=a$ and $r=b$ is self adjoint, we have $\varphi_1^{\dagger}(r)= \varphi_1(r)$ and $\Psi(r) = \varphi_1^2(r)$.}

\begin{equation}
\label{taboo_SDE_2D_long_time_regime_variance}
      \sigma^{*2} = \int_a^b \left( \frac{1}{r} \right)^2 \Psi(r) 2 \pi r dr  = 2 \pi  \int_a^b  \frac{ \varphi_1^2(r)}{r}  dr   \, . \\
\end{equation}

\noindent This quantity can be formally computed once $\varphi_1^2(r)$, given by Eq.(\ref{phi_1_polar_intermediate}), is normalized. 
Thus, even if the random variable $r_t$ remains non-trivial at stationarity, the angle $\theta^*_t$ of the taboo process in an annulus becomes a centered Gaussian variable whose evolution is given by,
\begin{equation}
\label{taboo_SDE_2D_polar_long_time_regime_theta}
	d \theta^*_t  = \sigma^* dW^{\theta}_t   \, .
\end{equation}

\noindent So, for large times, the angle $\theta^*_t$ of the taboo process in an annulus is equivalent of a Brownian motion on a circle of radius $r^* = 1/ \sigma^*$~\cite{ref_book_DelMoral,ref_Kundu}, 

\begin{equation}
\label{taboo_SDE_2D_long_time_regime_r}
	d \theta^*_t  =   \frac{d W_t}{r^*}  \qquad \mathrm{with}\qquad  r^* = \frac{1}{\sqrt{2 \pi  \int_a^b  \frac{ \varphi_1^2(r)}{r}  dr }} \, .
\end{equation}

\noindent This correspondence with Brownian motion on a circle has strong consequences since it implies that the known results for the Brownian motion on the circle apply for the taboo process in a annulus, provided that the radius $r^*$ is given by Eq.(\ref{taboo_SDE_2D_long_time_regime_r}). In particular, the transition probability density function of the $\theta^*_t$ process can be found in~\cite{ref_book_Bhattacharya} where diffusions on a circle are studied. In the same way, the recent results obtained by Kundu, Comtet and Majumdar~\cite{ref_Kundu} regarding the winding statistics of a Brownian particle on a ring are transferable to the taboo processes. Moreover, in~\cite{ref_Kundu}, the authors also consider the Brownian bridge on a circle, i.e. when the final angle $\theta^*_t$ of the particle is constrained to be a multiple of $2 \pi$. Furthers important analytical results were obtained that are also directly transferable to the taboo bridge on an annulus. We refer to the work~\cite{ref_Kundu} for these analytical results.\\

\begin{figure}[h]
\centering
\includegraphics[width=3.in,height=4.2in,angle=-90]{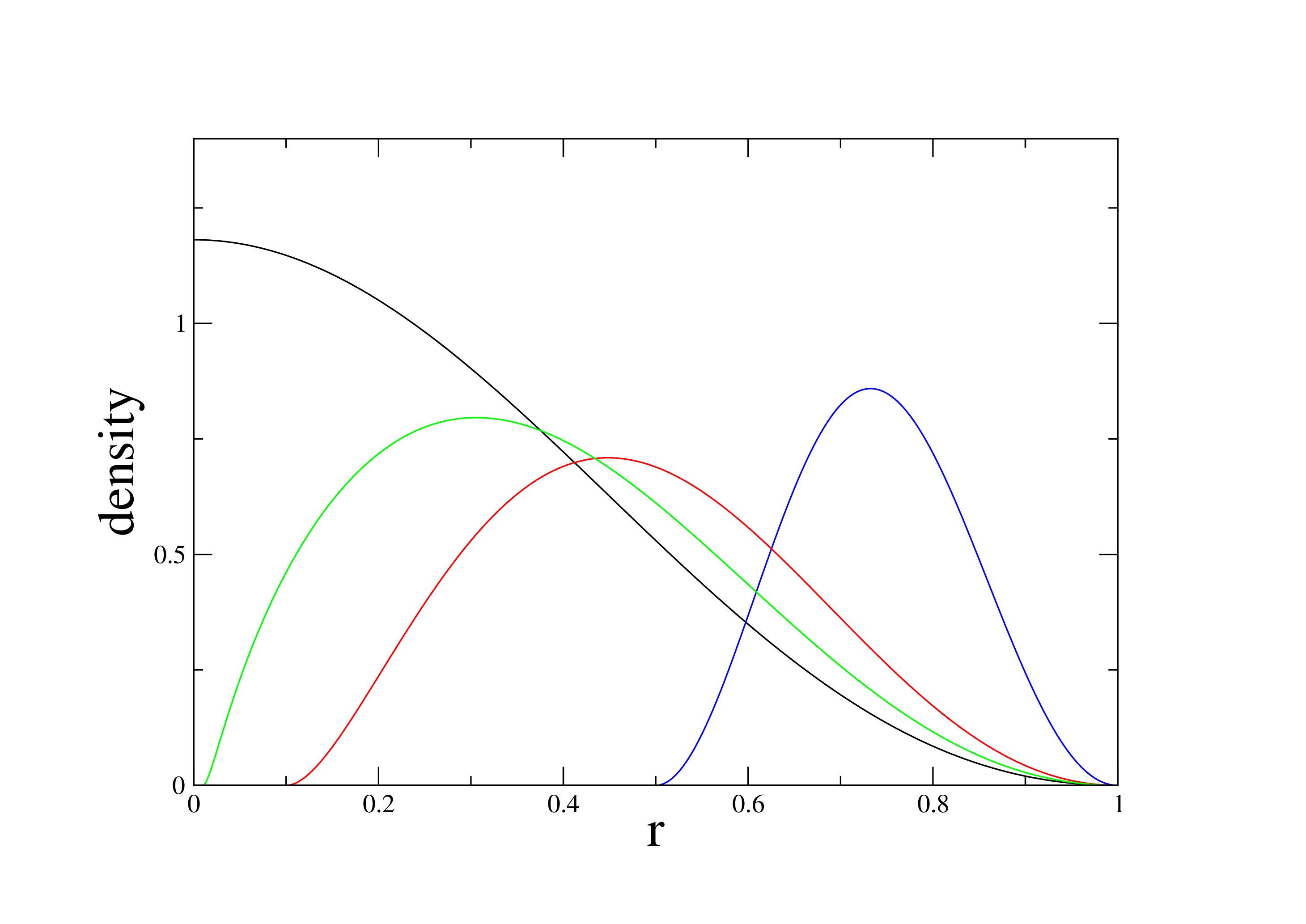}
\setlength{\abovecaptionskip}{15pt} 
\caption{Invariant density of the 2-dimension taboo process: black line corresponds to the disk. Blue, red and green lines correspond to the annulus with different inner radius: 0.5, 0.1 and 0.01.}
\label{fig8}
\end{figure}

Once the process has converged toward its stationary density, it spends most of its time around its mean value $\E[r^*_{\infty}]$. However, this position may be different from the center of the interval $[a,b]$ because of the curvature of the domain. Let us investigate this behavior more closely with a numerical example by considering a geometry where the annulus has a unit outer radius and an inner radius $100$ times smaller. We will compare the invariant density of this geometry with that of a unit disk in order to measure the influence of the small topological centered hole. To this aim, we need $\varphi_1(r)$ that we previously obtained in ~Eq.(\ref{phi_1_polar_intermediate}) to within a constant $C_0$. This constant is determined thanks to the normalization condition $\int_a^b \varphi_1^2(r) 2 \pi r dr =1$. We do not report this rather cumbersome expression of $C_0$, instead, we plot the invariant density of the taboo process in a annulus that we compare to the invariant density of the taboo process in the unit disk. The latter, established in appendix~\ref{appendix_6}, is given by the expression, 

\begin{equation}
\label{Invariant_density_2D_disk}
	 \Psi_{\mathrm{disk}}(r) = \frac{ J_0^2\left(z_1 r\right)}{\pi J_1^2(z_1)} \, ,
\end{equation}
\noindent where $z_1 \simeq 2.40483...$ is the first zero of the Bessel function $J_0(x)$. Figure~\ref{fig8}
shows the asymmetric behavior of the invariant density of the taboo process in an annulus for various values of the inner radius. Note that, when the inner radius shrinks to a point, the stationary density matches that of the taboo process in a disk. The mean value $\E[r^*_{\infty}]$ is given by the integral $\int_a^b r \Psi(r) 2 \pi r dr $ and is equal to $\E[r^*_{\infty}] \simeq 0.476$ with $b=1$ and $a=0.01$. A value that is fairly close to the mean value of the stationary position of the taboo process in a unit disk, $r^*_{\mathrm{disk}} \simeq 0.424$, given by Eq.(\ref{position_mean_value_disk_appendix_6}) (with $b=3$ and $a=1$, $\E[r^*_{\infty}] \simeq 1.997$ is plotted on Fig.~\ref{fig7}). By ignoring fluctuations, a rough estimate $\bar{r}^*$ of $\E[r^*_{\infty}]$ can be obtained, at least numerically, directly from Eq.(\ref{taboo_SDE_2D_polar_final}) by seeking the solution of $B(\bar{r}^*)+1/(2 \bar{r}^*) = 0$. Doing so, with $b=1$ and $a=0.01$, we find $\bar{r}^* \simeq 0.463$ in close agreement with the exact value $\E[r^*_{\infty}] \simeq 0.476$. \\

%%Once the long time regime is attained, the stochastic differential equation for $\theta_t$ becomes
%%\begin{equation}
%%\label{taboo_SDE_2D_polar_long_time_regime}
 %%     d \theta^*_t  = \frac{ d W^{\theta}_t}{r^*}  \, , \\
%%\end{equation}
%%
%%\noindent which is the stochastic differential equation for a Brownian motion on a circle of radius $r^*$~\cite{ref_book_DelMoral,ref_Kundu}. For large time, $r^*$ follows a probability density function, independent of $t$, which is precisely $\Psi(r)$. As a consequence, $\theta^*_t$ is a (continuous) sum of centered Gaussian random variables of variance $1/r^{*2}$. As such, it is also a centered Gaussian random variable, and its variance is given by,
%%
%%\begin{equation}
%%\label{taboo_SDE_2D_long_time_regime_variance}
%%      \sigma^{*2} = \int_a^b \left( \frac{1}{r} \right)^2 \Psi(r) 2 \pi r dr  = 2 \pi  \int_a^b  \frac{ \varphi_1^2(r)}{r}  dr   \, , \\
%%\end{equation}
%%
%%\noindent This number can be formally computed once $\varphi_1^2(r)$, given by Eq.(\ref{phi_1_polar_intermediate})), is normalized. So, for large times, the angle $\theta_t$ the taboo process in an annulus is equivalent of a Brownian motion on a circle of radius $r^* = 1/ \sigma^*$, i.e.,  
%%
%%\begin{equation}
%%\label{taboo_SDE_2D_long_time_regime_r}
%%	r^* = \frac{1}{\sqrt{2 \pi  \int_a^b  \frac{ \varphi_1^2(r)}{r}  dr }} \, .
%%\end{equation}

We end the study of the two-dimensional case with a general remark. Previously, we have seen that taboo processes are not the sole processes constrained to stay in a bounded domain forever. Other processes, such as generalized taboo processes or asymetric diffusion processes, have also this property. All these processes are ergodic and, as such, converge towards a stationary density independent of time. The reasoning that led us to the equations Eq.(\ref{taboo_SDE_2D_polar_long_time_regime_theta}) and Eq.(\ref{taboo_SDE_2D_long_time_regime_r}) based on this assumption therefore remains valid. We can conclude that in the large time limit, all processes constrained to stay in an annulus of radius $a$ and $b$ behave like a Brownian particle on a circle with an appropriate radius. More precisely, their angles $\theta_t$ satisfy the stochastic differential equation of a Brownian particle on a circle of radius $R$,

\begin{equation} 
\label{general_SDE_2D_polar_long_time_regime}
      d \theta_t  =   \frac{d W_t}{R}  \qquad \mathrm{with}\qquad  R = \frac{1}{\sqrt{2 \pi} \sqrt{ \int_a^b  \frac{ \Psi(r)}{r}  dr } } \\\
\end{equation}

\noindent where $\Psi(r)$ is, as usual, the stationary density of the process.

%%%%%%%%%%%%%%%%%%%%%%%%%%%%%%%%%%%%%%%%%%%%%%%%%%%%%%%%%%%%%%%%%%%%%%%%%%%%%%%%%%%%%%%%%%%%%%%%%%%%%%%%%%%%%%%%
%%																						%%
%%                 HIGHER DIMENSIONAL TABOO PROCESSES                                                         %%
%%																						%%
%%%%%%%%%%%%%%%%%%%%%%%%%%%%%%%%%%%%%%%%%%%%%%%%%%%%%%%%%%%%%%%%%%%%%%%%%%%%%%%%%%%%%%%%%%%%%%%%%%%%%%%%%%%%%%%%
\section{Higher dimensional taboo processes}
\label{sec_Taboo_higher_dimensions}

%%%%%%%%%%%%%%%%%%%%%%%%%%%%%%%%%%%%%%%%%%%%%%%%%%%%%%%%%%%%%%%%%%%%%
%%            THREE DIMENSIONAL TABOO PROCESSES                    %%
%%%%%%%%%%%%%%%%%%%%%%%%%%%%%%%%%%%%%%%%%%%%%%%%%%%%%%%%%%%%%%%%%%%%%
\subsection{Three dimensional taboo process}
\label{subsec_Taboo3D}
In this paragraph, we consider taboo processes in three-dimensional bounded domains. In particular, we study the behavior of the taboo process evolving inside a spherical shell, a non-convex region between two concentric spheres of radius $a$ and $b$,  $\mathcal{D}_A = {r \in R^3 : a < |r| < b}$. Taboo processes in other 3-dimensional geometries, like a cylinder, have been recently reported in~\cite{ref_Adorisio} under the name of "Brownian motion conditioned to stay inside a cylinder". For the spherical shell geometry, as in the two-dimensional annulus case, we start with a standard Brownian motion (here a three-dimensional one) and we seek the lowest eigenfunction of $-\mathcal{L} \varphi(r)  = -1/2 \Delta \varphi(r) = \lambda \varphi(r)$ ($\Delta$ being the 3-dimensional Laplacian) for the spherical shell with Dirichlet boundary conditions. The solution reads,

\begin{equation}
\label{phi_taboo_3D}
	\varphi(r)  = \frac{1}{\sqrt{2 \pi (b-a)}} \frac{\sin \left( \frac{\pi (r-a)}{b-a} \right)}{r} \, ,
\end{equation}

\noindent and in spherical coordinates the taboo process inside the spherical shell has generator,

\begin{equation}
\label{generator_taboo_3D}
      \begin{aligned}
	 \mathcal{L}. & = \frac{1}{2} \Delta . + \frac{{\bf{\nabla}} \varphi(r)}{\varphi(r)} {\bf{\nabla}} .\\  
			    & = \frac{1}{2}  \frac{{\partial}^2 .}{{\partial} r^2} + \frac{\pi}{b-a} \cot \left( \frac{\pi (r-a)}{b-a} \right)   \frac{{\partial} .}{{\partial} r}   + \frac{1}{2 r^2} \frac{{\partial}^2 .}{{\partial} \theta^2} + \frac{\cot \theta}{2 r^2} \frac{{\partial} .}{{\partial} \theta} + \frac{1}{2 r^2 \sin^2 \theta} \frac{{\partial}^2 .}{{\partial} \varphi^2}  \, .
      \end{aligned} 
\end{equation}
\noindent This generator corresponds to a diffusion process given by~\cite{ref_book_DelMoral}

\begin{equation}
\label{taboo_SDE_3D_polar_final}
  \left\{
      \begin{aligned}
      d r_t         & = \frac{\pi}{b-a} \cot \left( \frac{\pi (r_t-a)}{b-a} \right) dt + d W^{1}_t \\
      d \theta_t    & = \frac{\cot \theta_t}{2 {r_t}^2} \, dt + \frac{1}{r_t} \, d W^{2}_t \, , \\
      d \varphi_t   & = \frac{1}{r_t \,\sin \theta_t}\, d W^{3}_t \, , \\
      \end{aligned}  
    \right. \, 
\end{equation}

\noindent where  $W^{1}_t$, $W^{2}_t$ and $W^{3}_t$ are three independent standard Brownian motions. Observe that the first stochastic equation is independent of the two others. Moreover, since the process is ergodic, in the large time limit, it converges towards its invariant density (given by $\varphi^2(r)$). In this long time regime, we can easily calculate the exact average radial position,

%%rev   When this regime is reached, the last two stochastic equations would indicate that the process behaves like a Brownian on a sphere of radius $\E[1/r_t]$ with $t \to \infty$, if $\E[1/r^2_t]=\E[1/r_t]^2$. This is not strictly the case, but numerical examples with $b = 2a$ show that this approximation is reasonable for $a \ge 1$ since,
%%rev  \begin{eqnarray} 
%%rev  \hspace*{-1.2cm} \E \left[1/r^2_t\right] - \left(\E \left[1/r_t\right] \right)^2 & = -\frac{(\mathrm{Ci}(2 \pi )-\mathrm{Ci}(4 \pi )+\log (2))^2+2 \pi  (\mathrm{Si}(2 \pi )-\mathrm{Si}(4 \pi ))}{a^2} \nonumber \\
%%rev                                                                   & \simeq \frac{0.00708}{a^2} \, .
%%rev  \end{eqnarray}

%%rev  \noindent In any case, we can easily calculate the exact average radial position in the long time limit,

\begin{equation}
\label{r*_taboo_3D}
	\E[r] = \int_a^b r  \, \varphi^2(r) 4 \pi r^2 \, dr = \frac{a+b}{2} \, .
\end{equation}

\noindent Thus, the mean value of the radial position is the middle of the interval $[a,b]$, a result that is no longer valid when the dimension of space increases as it is outlined in the next section. When the radius $a$ shrinks to zero, the bounded domain is a sphere, and taboo processes evolving inside this convex region have been studied by Pinsky~\cite{ref_Pinsky} and Garbaczewski~\cite{ref_Garbaczewski}.

%%%%%%%%%%%%%%%%%%%%%%%%%%%%%%%%%%%%%%%%%%%%%%%%%%%%%%%%%%%%%%%%%%%%%
%%           HIGHER DIMENSIONAL TABOO PROCESSES                    %%
%%%%%%%%%%%%%%%%%%%%%%%%%%%%%%%%%%%%%%%%%%%%%%%%%%%%%%%%%%%%%%%%%%%%%
\subsection{Taboo processes in higher dimensions}
\label{subsec_Taboo_odd_dimensions}
Finally, we briefly consider the taboo process inside a spherical shell in $n$ dimensions, a region delimited by two concentric hyper-spheres of radius $a$ and $b$,  $\mathcal{D}_A = {r \in R^n : a < |r| < b}$. In order to obtain the expression of the drift term of the taboo process as well as its stationary density, we have to solve the eigenvalue equation for the operator $-1/2 \Delta_n$  ($\Delta_n$ being the $n$-dimensional Laplacian) with Dirichlet boundary conditions on $\partial \mathcal{D}_A$. Due to the spherical symmetry of the geometry, we seek a radial eigenfunction $\varphi(r)$ with a (lowest) eingenvalue $\lambda$ that satisfies 

\begin{equation}
     \frac{d^2 \varphi(r)}{dr^2} + \frac{n-1}{r} \frac{d \varphi(r)}{dr} = -2 \lambda \varphi(r)  \, .
\end{equation}

\noindent Since, $\lambda > 0$, the solution is expressed in terms of Bessel functions of the first and second kinds~\cite{ref_book_Polyanin} and the Dirichlet boundary conditions involve zeros of theses Bessel functions. 
We found numerically that as the dimension increases, the mean radial position of the taboo process gets closer to the outer hyper-sphere.\\

\noindent Note that the reflected Brownian motion behaves in the same way. Indeed, it is well-known that a Brownian motion trapped inside a bounded domain $\mathcal{D}$ by reflecting boundaries is ergodic, with a uniform stationary distribution $\Psi = 1/\mathrm{Vol(\mathcal{D})}$. The volume of a $d$-dimensional sphere being $\pi^{d/2} r^d/\Gamma(d/2+1)$, we get that the mean value of the radial position of the reflected Brownian motion in the d-dimensional spherical shell is

\begin{equation}
\label{r*_brownian_reflected_nD}
	\E[r_{ref}] = \int_a^b r  \, \frac{1} { \frac{\pi^{d/2}}{\Gamma(\frac{d}{2}+1)} (b^d -a^d)} \, \Omega_{d} \, r^{d-1} dr = \frac{d}{d+1} \frac{b^{d+1} -a^{d+1}}{b^d -a^d} \, .
\end{equation}

\noindent For $d=1$, we have $\E[r_{ref}] = \frac{a+b}{2}$ but for $d \gg 1$, $\E[r_{ref}] = b^-$, meaning that the reflected Brownian spends most of its time near the outer hyper-sphere as the number of dimensions increases.

%%%%%%%%%%%%%%%%%%%%%%%%%%%%%%%%%%%%%%%%%%%%%%%%%%%%%%%%%%%%%%%%%%%%%%%%%%%%%%%%%%%%%%%%%%%%%%%%%%%%%%%%%%%%%%%%
%%																						%%
%%                CONCLUSION                                                                                  %%
%%																						%%
%%%%%%%%%%%%%%%%%%%%%%%%%%%%%%%%%%%%%%%%%%%%%%%%%%%%%%%%%%%%%%%%%%%%%%%%%%%%%%%%%%%%%%%%%%%%%%%%%%%%%%%%%%%%%%%%
\section{Conclusion}
\label{sec_Conclusion}
In order to model the behavior of diffusive processes in confined geometries, we have introduced several generalizations of taboo processes, including the sweetest taboo process and the asymmetric diffusion process. We studied in detail these processes, in particular, in one dimension where we derived the stationary density as well as the speed of convergence towards this density. We have found analytical expressions for the speed of convergence for the generalized taboo processes and sharp bounds for the other processes. We also calculated in close form the mean exit time from an interval. In two dimensions, we studied the taboo process in a circular annulus. In the long time regime, based on the Gaussian behavior of the angle of the taboo process, we have highlighted a correspondence between the taboo process in an annulus and that of a Brownian motion on a circle. This correspondence allows us to transpose the deep results obtained for Brownian motions on a circle to taboo processes in a circular annulus. We conclude our study with an examination of the taboo process confined between two concentric hyper-spheres in high dimensions and  found out that the average radial position approaches the outer hyper-sphere. Besides, computer simulations of taboo processes are very easy since they are based on a modification of the drift term only and do not require the subtle boundary conditions of reflecting stochastic differential equations.
%% have been studied for several geometries in various dimensions. These processes are ergodic and, as such, converge toward a stationary density that one can easily obtain analytically, at least in simple geometries (typically with symmetries). 

%%%%%%%%%%%%%%%%%%%%%%%%%%%%%%%%%%%%%%%%%%%%%%%%%%%%%%%%%%%%%%%%%%%%%%%%%%%%%%%%%%%%%%%%%%%%%%%%%%%%%%%%%%%%%%%%
%%																						%%
%%                 ACKNOWLEDGMENTS                                                                            %%
%%																						%%
%%%%%%%%%%%%%%%%%%%%%%%%%%%%%%%%%%%%%%%%%%%%%%%%%%%%%%%%%%%%%%%%%%%%%%%%%%%%%%%%%%%%%%%%%%%%%%%%%%%%%%%%%%%%%%%%
\section{Acknowledgements}
% The author wish to thank .
The author is grateful to C. Dubois for suggesting the title and for her enlightening comments. 

%%%%%%%%%%%%%%%%%%%%%%%%%%%%%%%%%%%%%%%%%%%%%%%%%%%%%%%%%%%%%%%%%%%%%%%%%%%%%%%%%%%%%%%%%%%%%%%%%%%%%%%%%%%%%%%%
%%																						%%
%%                 APPENDIX                                                                                   %%
%%																						%%
%%%%%%%%%%%%%%%%%%%%%%%%%%%%%%%%%%%%%%%%%%%%%%%%%%%%%%%%%%%%%%%%%%%%%%%%%%%%%%%%%%%%%%%%%%%%%%%%%%%%%%%%%%%%%%%%
\appendix

%%%%%%%%%%%%%%%%%%%%%%%%%%%%%%%%%%%%%%%%%%%%%%%%%%%%%%
%%                 APPENDIX 1                       %%
%%%%%%%%%%%%%%%%%%%%%%%%%%%%%%%%%%%%%%%%%%%%%%%%%%%%%%
\section{Conditioned diffusion in an interval}
\label{appendix_1}
In this appendix, we give an informal derivation of the stochastic differential equation satisfies by the taboo process in an interval without referring to the more rigorous, but much more technical, work of Pinsky~\cite{ref_Pinsky}. The ingredients are results regarding the conditioning of a Brownian motion that can be found in~\cite{ref_book_Karlin}, the spectral representation of the transition density for a diffusion, again in~\cite{ref_book_Karlin} and the expression of the survival probability of a diffusion in an interval, subject to absorbing boundary conditions~\cite{ref_book_Redner}.\\ 
We briefly recall Doob's method for conditioning a process on its final state $T$: Consider a diffusion process in one dimension $\{X_t, 0 \le t \le T\}$ characterized by its drift $\mu$ and its variance $\sigma^2$. The diffusion $X_t$ is driven by, 

\begin{equation}
\label{diffusion_appendix_1}
  dX_t = \mu dt+ \sigma dW_t ,
\end{equation}

\noindent where $W_t$ is a standard Brownian motion. $\mathcal{L}. =    \mu  \frac{d .}{d x}  + \frac{\sigma^2}{2}  \frac{d^2 .}{d x^2}$ is the generator of the diffusion. Now, let $\{X^*_t, 0 \le t \le T\}$ be the process conditioned on its state at time $T$. Then, the drift $\mu^*(x,t,T)$ and the variance $\sigma^{*2}(x,t,T)$ of the constrained process are given by~\cite{ref_book_Karlin} :

\begin{equation}
\label{variance_appendix_1}
  \left\{
      \begin{aligned}
        \sigma^*(x,t,T) & = \sigma             \\
        \mu^*(x,t,T)    & = \mu + \frac{\partial \pi(x,t,T)/\partial x}{\pi(x,t,T)} \sigma^2 \, , \\
      \end{aligned}
    \right.
\end{equation}

\noindent where $\pi(x,t,T)$ is the probability that from the state value $x$ at time $t$, the sample path of $X_t$ satisfies the constraint at time $T$.  Next, consider an interval $[-L,L]$ with absorbing conditions on the boundaries. The process conditioned to remain in the interval up to time $T$ corresponds to realizations that have not hit the boundaries, i.e. to trajectories that have survive up to time $T$. The probability $\pi(x,t,T)$ is thus the survival probability till time $T$ and $\lim_{T \to \infty} \pi(x,t,T) = \pi(x,t)$ is the asymptotic survival probability. The drift and the variance of taboo process correspond to the limit when $T \to \infty$ since the taboo process survives forever. Besides, the survival probability is related to the transition density by the relationship

\begin{equation}
\label{survival_and_probability_density_appendix_1}
 	 \pi(x,t,T) = \int_{-L}^{L} p(x,y,T-t) \, dy  \, ,
\end{equation}

\noindent where $p(x,y,T-t) dy$ is the probability that the process is at $y$ in the interval $dy$ at time $T$ given that it started at time $t$ at the position $x$. It is well known that the transition density has a spectral representation~\cite{ref_book_Karlin}

\begin{equation}
\label{spectral_probability_density_appendix_1}
 	 p(x,y,T-t) = m(y) \sum\limits_{n=1}^{\infty} e^{-\lambda_n (T-t)} \varphi_n(x)  \varphi_n(y) \pi_n \, ,
\end{equation}

\noindent where $m(x) = e^{-2 \mu x/\sigma^2}/\sigma^2$ is the speed density function defined in appendix~\ref{appendix_3} (Eqs.(\ref{scale_function_and_more})), and $\varphi_n(x)$ the eignefunctions of the equation $- \mathcal{L}\varphi_n(x) = \lambda_n \varphi_n(x) $ 
%%\begin{equation}
%%\label{eigenfunction_appendix_1}
%%	 \frac{\sigma^2}{2} \frac{d^2 \varphi_n(x)}{d x^2} + \mu  \frac{d \varphi_n(x)}{d x}  = -\lambda_n  \varphi_n(x)
%%\end{equation}
%%\noindent 
with Dirichlet boundary conditions. $\varphi_n(x)=e^{-\mu x/\sigma^2} \cos\left(\frac{n \pi x}{2 L}\right)$, and  
$\lambda_n=\frac{\mu^2}{2 \sigma^2} + \frac{\sigma^2}{2} \left( \frac{n \pi}{2 L} \right)^2$ is the $n$-th eigenvalue. $\pi_n$ are coefficients that can be easily calculated: $1/\pi_n = \int_{-L}^{L}  m(y)\varphi_n(y)\, dy$. The detailed description of these objects is not important. What is crucial is that the eigenvalue spectrum has a smaller positive eigenvalue $\lambda_1$ and in the long time limit, only the lowest eigenmode remains. In this regime, we thus have,

\begin{equation}
\label{survival_appendix_1}
	\pi(x,t) = \lim_{T \to \infty} \int_{-L}^{L} p(x,y,T-t) \, dy \sim  \varphi_1(x)   \pi_1 e^{-\lambda_1 (T-t)}  \int_{-L}^{L} m(y) \varphi_1(y)  \, dy \,
\end{equation}

\noindent and the drift of the taboo process is given by

\begin{equation}
\label{drift_taboo_appendix_1}
	\mu^*(x,t)  = \lim_{T \to \infty} \mu^*(x,t,T)  = \lim_{T \to \infty} \left[ \mu + \frac{\partial \pi(x,t,T)/\partial x}{\pi(x,t,T)} \sigma^2 \right] = \mu + \frac{d\varphi_1(x)/d x}{\varphi_1(x)} \sigma^2 \, .
\end{equation}

\noindent This results holds as long as the survival probability can be represented as an eigenfunction expansion of decaying exponential in time, where only the lowest eigenmode remains in the long time limit. 
%In one dimension this is indeed the case when the diffusion coefficient $\sigma^2(x)$ is continuous and strictly positive and the drift coefficient $\mu(x)$ is continuous, both on the interval $[-L,L]$ with Dirichlet boundary conditions.
In one dimension this is indeed the case for a diffusion on the interval $[-L,L]$ with Dirichlet boundary conditions~\cite{ref_book_Karlin,ref_book_Risken}. In higher dimensions, the preceding eigenfunction expansion remains valid if we further assume that the generator $\mathcal{L} . =  \sum_{i=1}^n  \mu_i(\bm{x}) \frac{\partial .}{\partial x_i}  + \frac{1}{2} \sum_{i=1}^n \sum_{j=1}^n \left( \bm{\sigma}(\bm{x}) \bm{\sigma}(\bm{x})^\mathsf{T} \right)_{i,j} \frac{\partial^2 .}{\partial x_i \partial x_j}$ of the $n$-dimensional diffusion is self-adjoint~\cite{ref_book_Risken}. In particular, this is the case for the $n$-dimensional Brownian motion, with $\mu_i(\bm{x}) = \bm{0}$  and $\bm{\sigma}(\bm{x}) \bm{\sigma}(\bm{x})^\mathsf{T} = \sigma^2 I_n$, where $I_n$ is the identity matrix.

%%\begin{equation}
%%\label{generator_appendix_1}
%%	\mathcal{L} . =  \sum_{i=1}^n  \mu_i(\bm{x}) \frac{\partial .}{\partial x_i}  + \frac{1}{2} \sum_{i=1}^n \sum_{j=1}^n \left( \bm%%{\sigma}(\bm{x}) \bm{\sigma}(\bm{x})^\mathsf{T} \right)_{i,j} \frac{\partial^2 .}{\partial x_i \partial x_j} \, , 
%%\end{equation}

%%\begin{equation}
%%\label{survival_appendix_1}
%% 	 \pi(x,t) \sim \cos\left(\frac{\pi x}{2 L}\right) \exp{\left[- \left(\frac{\mu^2}{2 \sigma^2} + \frac{\pi^2 \sigma^2}{8 L^2} \right) (T-t) \right]} \, .
%%\end{equation}

%%%%%%%%%%%%%%%%%%%%%%%%%%%%%%%%%%%%%%%%%%%%%%%%%%%%%%
%%                 APPENDIX 2                       %%
%%%%%%%%%%%%%%%%%%%%%%%%%%%%%%%%%%%%%%%%%%%%%%%%%%%%%%
\section{Eigenfunctions and stationary density in one dimension}
\label{appendix_2}
In this appendix, we consider a one-dimensional diffusion operator $\mathcal{L}$ and its adjoint $\mathcal{L}^{\dagger}$ on an interval. Let $\varphi_1(x)$ be the lowest eigenfunction of $-\mathcal{L}$ with Dirichlet boundary conditions on the border of the interval, $\lambda_1$ its eigenvalue, and $\Psi(x)$ is the stationary density of the associated taboo process. We state that $\varphi_1^{\dagger}(x) = \Psi(x)/\varphi_1(x)$ is the corresponding lowest eigenfunction of $-\mathcal{L}^{\dagger}$ (with the same boundary conditions) and that  $\lambda_1$ is also its eigenvalue.\\

\noindent Consider a one-dimensional diffusion operator $\mathcal{L}$ with parameters $\mu(x)$ and $\sigma(x)$, 
\begin{equation}
\label{generator_scalar_appendix_2} 
	\mathcal{L}.  =    \mu(x)  \frac{d .}{dx}  + \frac{\sigma(x)^2}{2}  \frac{d^2 .}{d x^2} \, , 
\end{equation}
\noindent and
\begin{equation}
\label{generator_adjoint_scalar_appendix_2}
	\mathcal{L}^{\dagger} . =   - \frac{d (\mu(x) .)}{d x}  + \frac{1}{2} \frac{d^2 (\sigma(x)^2 .)}{d x^2} \, 
\end{equation}
\noindent its formal adjoint. The corresponding taboo process has parameters, $\mu^*(x)$ and $\sigma^*(x)^{2}$ given by,

\begin{equation}
\label{constrained_drift_variance_appendix_2}
  \left\{
      \begin{aligned}
        \sigma^{*}(x) & = \sigma(x)   \\
        \mu^*(x) & = \mu(x)  + \sigma(x)^2 \frac{\varphi_1'(x)}{\varphi_1(x)} \, , \\
      \end{aligned}
    \right.
\end{equation}

\noindent where $\varphi_1(x)$ is the lowest eigenfunction of $-\mathcal{L}$ with Dirichlet boundary conditions. This eigenfunction obeys the equation

\begin{equation}
\label{eq_first_eigenfunction_1D_appendix_2}
	\mu(x)  \frac{d \varphi_1(x)}{d x} + \frac{\sigma(x)^2}{2} \frac{d^2 \varphi_1(x)}{d x^2} = -\lambda_1 \varphi_1(x) .
\end{equation}

\noindent The stationary density $\Psi(x)$ of the associated taboo process satisfies the Fokker-Planck equation,

\begin{equation}
\label{eq_fokker_planck_density_1D_appendix_2}
	- \frac{d ( \mu^*(x) \Psi(x))}{d x}  + \frac{1}{2} \frac{d^2 (\sigma^*(x)^{2} \Psi(x))}{d x^2} = -\frac{d J(x)}{d x} = 0 \, ,
\end{equation}

\noindent where 
\begin{equation}
\label{eq_current_1D_appendix_2}
	J(x) := \left[\mu^*(x) - \frac{1}{2} \frac{d }{d x} \sigma^*(x)^{2} \right] \Psi(x)
\end{equation}

\noindent is the probability current of the taboo process. The preceding equation indicates that the probability current is constant. Moreover, for the taboo process, the two boundaries are entrance boundaries. Therefore, the probability current of the taboo process vanishes at the boundaries, and the current must be zero anywhere within the interval. Then, since $J(x) = 0$,

\begin{equation}
\label{eq_current_zero_1D_appendix_2}
	\left[\mu^*(x) - \frac{1}{2} \frac{d }{d x} \sigma^*(x)^{2} \right] \Psi(x) = 0.
\end{equation}

\noindent An immediate integration yields to,

\begin{equation}
\label{Invariant_density_1D_appendix_2_aux}
	 \Psi(x) = \frac{N}{\sigma^*(x)^{2}} e^{\int^x \frac{2 \mu^*(\xi)}{\sigma^*(\xi)^{2}} d\xi}
\end{equation}

\noindent where $N$ is a constant that ensures the normalization of the density $\Psi(x)$. Inserting the expressions of the parameters of $\mu^*(x)$ and $\sigma^*(x)$ given by Eq.(\ref{constrained_drift_variance_appendix_2}) into the preceding equation leads to,

\begin{equation}
\label{Invariant_density_1D_appendix_2}
	 \Psi(x) = N \frac{\varphi_1(x)^2}{\sigma(x)^{2}} e^{\int^x \frac{2 \mu(\xi)}{\sigma(\xi)^{2}} d\xi} \, .
\end{equation}

\noindent At the beginning of the paragraph, we claim that $\varphi_1^{\dagger}(x)  = \Psi(x)/\varphi_1(x)$ is lowest eigenfunction of $-\mathcal{L}^{\dagger}$ and $\lambda_1$ its eigenvalue. The proof now follows a direct calculation. Plugging,

\begin{equation}
\label{eq_first_eigenfunction_adjoint_1D_appendix_2}
	 \varphi_1^{\dagger}(x) = \frac{\Psi(x)}{\varphi_1(x)} = N \frac{\varphi_1(x)}{\sigma(x)^{2}} e^{\int^x \frac{2 \mu(\xi)}{\sigma(\xi)^{2}} d\xi} \, ,
\end{equation}

\noindent into the adjoint $\mathcal{L}^{\dagger}$ gives

\begin{equation}
\label{eq_adjoint_1D_appendix_2}      
	\begin{aligned}
	 \mathcal{L}^{\dagger} \varphi_1^{\dagger}(x) & = - \frac{d \left[ \mu(x) \scriptstyle{\times} \displaystyle N \frac{\varphi_1(x)}{\sigma(x)^{2}} e^{\int^x \frac{2 \mu(\xi)}{\sigma(\xi)^{2}} d\xi} \right]}{d x}  + \frac{1}{2} \frac{d^2 \left[ \sigma(x)^2 \scriptstyle{\times} \displaystyle  N \frac{\varphi_1(x)}{\sigma(x)^{2}} e^{\int^x \frac{2 \mu(\xi)}{\sigma(\xi)^{2}} d\xi}\right]}{d x^2} \\
      & = N \frac{ e^{\int^x \frac{\mu(\xi)}{\sigma(\xi)^{2}} d\xi} } {\sigma(x)^{2}} \scriptstyle{\times} \displaystyle \underbrace{ \left[  \mu(x)\frac{d \varphi_1(x)}{d x} + \frac{\sigma(x)^2}{2} \frac{d \varphi_1(x)}{d x} \right] }_{-\lambda_1 \varphi_1(x) \text{ by virtue of Eq.} \eqref{eq_first_eigenfunction_1D_appendix_2} }    \\
      & =  - \lambda_1 \varphi_1^{\dagger}(x)  \, ,
 	\end{aligned}
\end{equation}

\noindent which is the announced result. Remark that in section~\ref{sec_Taboo1D}, we obtained this result by a direct calculation of both eigenvalues, but in the particular case where the infinitesimal drift and variance were constant.

%%%%%%%%%%%%%%%%%%%%%%%%%%%%%%%%%%%%%%%%%%%%%%%%%%%%%%
%%                 APPENDIX 3                       %%
%%%%%%%%%%%%%%%%%%%%%%%%%%%%%%%%%%%%%%%%%%%%%%%%%%%%%%
\section{Entrance boundary classification and stationary density}
\label{appendix_3}
Throughout this article, we make a repeated use of the concept of entrance boundary and stationary density. 
In this appendix, we review the elements that characterize such a boundary for a one-dimensional diffusion process on the state space $[l,r]$ with drift and diffusion coefficients $\mu(x)$ and $\sigma^2(x)$. When the two boundaries are entrance, we also give the expression of the stationary density. To this aim, we follow the excellent presentation given in the book of Karlin and Taylor~\cite{ref_book_Karlin} where all type of boundaries are examined.

\subsection{Entrance boundary}
Recall that an entrance boundary is a boundary that cannot be reached from the interior of the state space. 
Karlin and Taylor's book provides a neat criterion for a border to be an entrance boundary, but it requires several auxiliary functions. We introduce them now. First of all, let $S(x)$ be the scale function and $m(x)$ the speed density,
\begin{equation}
\label{scale_function_and_more}
  \left\{
      \begin{aligned}
        S(x) & = \int_{x_0}^x s(\eta)  d\eta  \qquad \mathrm{with~~} s(x) = \exp \left(- \int_{x_0}^{x} \frac{2 \mu(\xi)}{\sigma^2(\xi)} d\xi \right)                 \\
        m(x) & = \frac{1}{\sigma^2(x) s(x)}  \, , \\
      \end{aligned}
    \right.
\end{equation}

\noindent where $x_0$ is an arbitrary fixed point (of no relevance) in the interior of the state space (in practice, the point $x_0$ is selected so as to make the calculations simplest. For the taboo process it is the middle of the interval $[l,r]$). Next, we introduce the scale measure $S(l,x]$ ($l$ being the left boundary) and the speed measure $M(l,x]$,
\begin{equation}
\label{scale_and_speed_measure}
  \left\{
      \begin{aligned}
        S(l,x] & =  \int_l^x s(\eta)  d\eta  \\
        M(l,x] & =  \int_l^x m(\eta)  d\eta \, ,\\
      \end{aligned}
    \right.
\end{equation}

\noindent and finally a last quantity,
\begin{equation}
\label{def_N_l}
        N(l)  =  \int_l^x M(l,\xi]  s(\xi) d\xi \, .
\end{equation}
\noindent A precise meaning and a deep understanding of these functions are given in Karlin and Taylor's book. The criterion for the boundary $l$ to be an entrance boundary is the following:

\begin{equation}
\label{theorem_entrance}
   l \mathrm{~is ~an ~entrance ~boundary ~if~} S(l,x] = \infty \mathrm{~~and~~} N(l) < \infty \, .
\end{equation}

\subsection{Stationary density}
Again, we consider the one-dimensional diffusion process on the sate space $[l,r]$ with drift and diffusion coefficients $\mu(x)$ and $\sigma^2(x)$. We also assume that the stochastic process cannot escape the boundaries. 
Such a process is ergodic, meaning that the distribution of the diffusion converges to a (unique) stationary distribution $\Psi(x)$ as $t \to \infty$. This invariant distribution is related to the scale function and speed density (introduced in the preceding paragrah) through the formula~\cite{ref_book_Karlin},

\begin{equation}
      \Psi(x) =  m(x)\left[ C_1 S(x) + C_2\right],
\end{equation}

\noindent where the two constants $C_1$ and $C_2$ are determined such that $\Psi(x)$ is positive on $[l,r]$ and normalized $\int_l^r \Psi(x) dx =1$. A fine remark due to Karlin and Taylor simplifies the calculation of the stationary density when the two endpoints of the interval are entrance boundaries. Indeed, in their boundary classification, they noticed that at the right boundary, $\lim_{x \to r} S(x) = - \infty$. Since $\Psi(x)$ must be positive throughout the interval $[l,r]$, this implies that $C_1 = 0$. $C_2$ being determined by the normalization, we finally get,
 
\begin{equation}
\label{stationary_density_simplified}
      \Psi(x) =  \frac{m(x)}{\int_0^R m(\eta) d\eta} = \frac{1}{\sigma^2(x) s(x) \int_0^R \frac{d\eta}{\sigma^2(\eta) s(\eta)} } .
\end{equation} 
An expression that do not required the knowledge of the smallest eigenfunction of the $-\mathcal{L}$ operator as in Pinsky's approach.

\subsection{Taboo process stationary density}
As a direct application, we calculate the stationary density of a taboo process in the interval $[-L,L]$. For such a process, from Eq.(\ref{taboo_SDE_1D_interval}) we have $\mu(x)= - \pi \sigma^2/(2L) \tan \left( \pi x/(2L)  \right)$ and $\sigma(x)=\sigma$, therefore from Eqs.(\ref{scale_function_and_more}) we get,
\begin{equation}
  \left\{
      \begin{aligned}
        s(x) & = 1/\cos^2 \left( \frac{\pi x}{2 L} \right) \\
	   m(x) & = 1/s(x)=\cos^2 \left( \frac{\pi x}{2L} \right) \, . \\
      \end{aligned} 
    \right.
\end{equation}

\noindent Reporting these expressions in Eq.(\ref{stationary_density_simplified}) leads immediately to
\begin{equation}
      \Psi(x) =  \frac{1}{L} \cos^2 \left( \frac{\pi x}{2L} \right),
\end{equation}
\noindent which is the invariant density Eq.(\ref{invariant_measure_1_dimension}) obtained by Pinsky's procedure.

%%%%%%%%%%%%%%%%%%%%%%%%%%%%%%%%%%%%%%%%%%%%%%%%%%%%%% 
%%                 APPENDIX 4                       %%
%%%%%%%%%%%%%%%%%%%%%%%%%%%%%%%%%%%%%%%%%%%%%%%%%%%%%%
\section{Spectral gap theorems for ergodic diffusion processes}
\label{appendix_4}
In this appendix, for sake of completeness, we recall some recent useful theorems regarding the spectral gap of one-dimensional diffusion processes. Reference to this section is Pinsky's work~\cite{ref_Pinsky_gap} where proofs are given. Again, we consider a diffusion process $X_t$ on an interval $[-L,L]$ with generator,
\begin{equation}
\label{generator_ergodic}
	\mathcal{L}. =   \frac{\sigma^2}{2}   \frac{d^2 .}{d x^2}  + \mu(x)  \frac{d .}{d x}  \,  
\end{equation}
where $\mu(x)$ is piecewise $C^1$ on $[-L,L]$. We assume that the diffusion is ergodic, meaning that the distribution of the process converges to a stationary density as $t \to \infty$. This condition is fulfill when both boundaries, $-L$ and $L$, are entrance boundaries. These boundary conditions in turn impose integral conditions on $\mu(x)$ (given in section~\ref{sec_asymmetric_diffusion}). Moreover, let us note $\mu_T(x)$, the drift corresponding to the taboo process in the interval $[-L,L]$, i.e.
\begin{equation}
\label{drift_taboo}
	\mu_T(x) = - \frac{\pi \sigma^2}{2L} \tan \left( \frac{\pi x}{2L} \right) \, . 
\end{equation}
Under Dirichlet boundary conditions, it is well known that the spectra of the $-\mathcal{L}$ operator has discrete eigenvalues $\lambda_n(\mu)$ that can be arranged in non-decreasing order:
\begin{equation}
\label{eigenvalues_b}
	0 =  \lambda_0(\mu) < \lambda_1(\mu) \le \lambda_2(\mu) \le \dots
\end{equation}
and the spectral gap, i.e. the distance between the first two eigenvalues, is thus equal to $\lambda_1(\mu)$. Furthermore, assume that $\mu(x)$ is anti-symmetric then, according to Pinsky~\cite{ref_Pinsky_gap} we have the following results:

\begin{enumerate}
\item If $\mu(x) \ge 0$ on $[-L,0]$  and  $\mu(x) \le 0$ on $[0,L]$ then~:
\begin{equation}
\label{thm_reflected}
	 \lambda_1(\mu) \ge \frac{\sigma^2}{2}\left(\frac{\pi}{2L}\right)^2 \, ,
\end{equation}
with equality only in the case of the reflected Brownian motion on $[-L,L]$, i.e. when $\mu(x) = 0$.
\item If $\mu(x) \le 0$ on $[-L,0]$  and  $\mu(x) \ge 0$ on $[0,L]$ then~:
\begin{equation}
%%\label{thm_reflected}
	 \lambda_1(\mu) \le \frac{\sigma^2}{2}\left(\frac{\pi}{2L}\right)^2 \, ,
\end{equation}
with equality only in the case of the reflected Brownian motion on $[-L,L]$, i.e. when $\mu(x) = 0$.
\item If $\mu(x) \ge \mu_T(x)$ on $[-L,0]$  and  $\mu(x) \le \mu_T(x)$ on $[0,L]$ then~:
\begin{equation}
\label{thm_comparison}
   	\lambda_1(\mu) \ge \frac{3 \sigma^2}{2}\left(\frac{\pi}{2L}\right)^2 \, ,
\end{equation} 
with equality only in the case of the taboo process on $[-L,L]$, i.e. when $\mu(x) = \mu_T(x)$.
\item If $\mu(x) \le \mu_T(x)$ on $[-L,0]$  and  $\mu(x) \ge \mu_T(x)$ on $[0,L]$ then~:
\begin{equation}
%%\label{thm_comparison}
   	\lambda_1(\mu) \le \frac{3 \sigma^2}{2}\left(\frac{\pi}{2L}\right)^2 \, ,
\end{equation} 
with equality only in the case of the taboo process on $[-L,L]$, i.e. when $\mu(x) = \mu_T(x)$.
\end{enumerate}

\noindent We also state a last more intuitive result. Consider two different drifts $\mu(x)$ and $\tilde{\mu}(x)$ of two ergodic diffusion processes in an interval. If the drift $\tilde{\mu}(x)$ is larger than the drift $\mu(x)$ (which physically corresponds to more repulsive walls), then the diffusion associated with the drift $\tilde{\mu}(x)$ converges towards its stationary density more rapidly than that with the drift $\mu(x)$. Assume again that $\mu(x)$ is anti-symmetric, then (Pinsky~\cite{ref_Pinsky_gap})

\begin{enumerate}
\item If $\tilde{\mu}(x) \ge \mu(x)$ on $[-L,0]$  and  $\tilde{\mu}(x) \le \mu(x)$ on $[0,L]$ then~:
\begin{equation}
\label{thm_comparison_spectral_gap}
   	\lambda_1(\tilde{\mu}) >  \lambda_1(\mu)
\end{equation}
\item  If $\tilde{\mu}(x) \le \mu(x)$ on $[-L,0]$  and  $\tilde{\mu}(x) \ge \mu(x)$ on $[0,L]$ then~:
\begin{equation}
   	\lambda_1(\tilde{\mu}) <  \lambda_1(\mu) \,  .
\end{equation}
\end{enumerate}

%%%%%%%%%%%%%%%%%%%%%%%%%%%%%%%%%%%%%%%%%%%%%%%%%%%%%%
%%                 APPENDIX 5                       %%
%%%%%%%%%%%%%%%%%%%%%%%%%%%%%%%%%%%%%%%%%%%%%%%%%%%%%%
\section{Spectral gap: correspondence between diffusion and Schr\"odinger operators in one dimension}
\label{appendix_5}
In this appendix, we highlight the correspondence between an ergodic diffusion operator on an interval and its associated Schr\"odinger operator. 
In particular, we show that a constant potential is the corresponding Schr\"odinger operator of the taboo process. Since a constant potential plays no role for the spectral gap (it shifts all the eigenvalues by a constant), an immediate consequence is that the spectral gap of the taboo process and that of a Schr\"odinger operator without potential are identical. It is precisely this correspondence that allowed us in section~\ref{sec_Taboo1D} to calculate the spectral gap of the taboo process. We will detail our point now.\\
 
\noindent Again, consider a one-dimensional diffusion operator $\mathcal{L}$ with parameters $\mu(x)$ and $\sigma(x)$, and assume that the parameters are such that both boundaries are entrance boundaries,

\begin{equation}
\label{generator_scalar_appendix_5}
	\mathcal{L}.  =   \frac{\sigma(x)^2}{2}  \frac{d^2 .}{d x^2} +  \mu(x)  \frac{d .}{dx}  \, . 
\end{equation}

\noindent It is well known that the correspondence between a diffusion process with entrance boundaries and a Schr\"odinger operator  $\mathcal{H}$ with absorbing walls (i.e. Dirichlet boundary conditions) is given by~\cite{ref_book_Risken}

\begin{equation}
\label{schrodinger_operator_appendix_5}
	\mathcal{H}.  =    \frac{d}{d x}  \frac{\sigma(x)^2}{2} \frac{d .}{d x} - V(x)  \, . 
\end{equation}

\noindent with the potential

\begin{equation}
\label{potential_appendix_5}
	V(x)  =  \frac{1}{2\sigma(x)^2}  \left(  \frac{1}{2} \frac{d\sigma(x)^2}{d x} - \mu(x)  \right)^2  + \frac{1}{2} \frac{d\mu(x)}{d x}  - \frac{1}{4}  \frac{d^2 \sigma(x)^2}{d x^2}       \, . 
\end{equation}

\noindent Plugging the taboo process parameters, $\mu_T(x)= - \frac{\pi \sigma^2}{2L} \tan \left( \frac{\pi x}{2L}  \right) $ and $\sigma_T(x)= \sigma$ into Eq.(\ref{potential_appendix_5}) and reporting the result into  Eq.(\ref{schrodinger_operator_appendix_5}) gives

\begin{equation}
\label{schrodinger_operator_taboo_appendix_5}
	\mathcal{H_T}.  =  \frac{\sigma^2}{2}  \frac{d^2 .}{d x^2}  -  \frac{\pi^2 \sigma^2}{8 L^2} \, . 
\end{equation}

\noindent The constant term $-\pi^2 \sigma^2/8 L^2$ (which corresponds to the ground state energy of the system) shifts all the eigenvalues of the same value, and thus does not act on the spectral gap (which is the difference between the two lowest eigenvalues). Therefore, we have the following result. The taboo process

\begin{equation}
\label{generator_taboo_appendix_5}
	\mathcal{L_T}.  =   \frac{\sigma^2}{2}  \frac{d^2 .}{d x^2} - \frac{\pi \sigma^2}{2L} \tan \left( \frac{\pi x}{2L}  \right)  \frac{d .}{dx}   
\end{equation}

\noindent and the Schr\"odinger operator with Dirichlet boundary conditions

\begin{equation}
\label{schrodinger_operator_scalar_simple_appendix_5}
	\mathcal{H}.  =  \frac{\sigma^2}{2}  \frac{d^2 .}{d x^2} 
\end{equation}

\noindent have the same spectral gap.\\

\noindent Note that in section~\ref{sec_Taboo1D} we evaluated the spectral gap with the following diffusion operator, 

\begin{equation}
\label{generator_scalar_constant_appendix_5}
	\mathcal{L}.  =   \frac{\sigma^2}{2}  \frac{d^2 .}{d x^2} +  \mu  \frac{d .}{dx}  \, . 
\end{equation}

\noindent The additional constant drift term $\mu$ can be remove thanks to the same procedure, leading to a Schr\"odinger operator with a constant potential that can be (again) shifted to zero. In fact, in section~\ref{sec_Taboo1D} we could have worked directly with the simpler Eq.(\ref{schrodinger_operator_scalar_simple_appendix_5}) instead of Eq.(\ref{generator_scalar_constant_appendix_5}) to obtain the spectral gap of the taboo process.

%%%%%%%%%%%%%%%%%%%%%%%%%%%%%%%%%%%%%%%%%%%%%%%%%%%%%%
%%                 APPENDIX 6                       %%
%%%%%%%%%%%%%%%%%%%%%%%%%%%%%%%%%%%%%%%%%%%%%%%%%%%%%%
\section{Stationary density in the disk}
\label{appendix_6}
In this last appendix, for sake of completeness, we derive the stationary density of the taboo process in a disk of radius $R$. The operator $-1/2 \Delta_2$ with Dirichlet conditions on $r=R$ being self adjoint, we can choose the eigenfunctions such that $\varphi_1^{\dagger}(r) = \varphi_1(r)$. Thus, we have that the invariant density is $\Psi_{\mathrm{disk}}(r) = \varphi_1^2(r)$. The solution of the eigenvalue equation $-1/2 \Delta_2 \varphi_1(r)= \lambda_1 \varphi_1(r)$ is given in term of Bessel functions and two constants $C_2$ and $C_3$ (Eq.(\ref{taboo_2D_phi_1_start})),
\begin{equation}
      \varphi_1(r) = C_2 J_0(\sqrt{2 \lambda_1} r) +  C_3 Y_0(\sqrt{2 \lambda_1} r).
\end{equation}
\noindent The divergence of $Y_0(x)$ at the origin imposes that $C_3 = 0$, and the boundary condition $\varphi_1(R) = 0$ implies that $\sqrt{2 \lambda_1} = z_1/R$ where $z_1 \simeq  2.40483...$ is the first zero of the Bessel function $J_0(x)$. The constant $C_2$ is then determined thanks to the density normalization 
\begin{equation}
	1 = \int_0^R \Psi_{\mathrm{disk}}(r) 2 \pi r dr = \int_0^R  C_2^2 J_0^2 \left( z_1 \frac{r}{R} \right) 2 \pi rdr = C_2^2 \pi R^2 J_1^2(z_1) \, ,
\end{equation}
\noindent therefore

\begin{equation}
\label{invariant_density_disk_appendix_6}
  \left\{
      \begin{aligned}
       \varphi_1(r) & = \frac{J_0 \left( z_1 \frac{r}{R} \right)}{\sqrt{\pi} R J_1(z_1)}  \\
       \Psi_{\mathrm{disk}}(r)   & = \frac{J_0^2\left( z_1 \frac{r}{R} \right)}{\pi R^2 J_1^2(z_1)}          \, . \\
      \end{aligned}
    \right.
\end{equation}

\noindent The mean value of the stationary position follows immediately,
\begin{equation}
\label{position_mean_value_disk_appendix_6}
	\int_0^R r \, \Psi_{\mathrm{disk}}(r) 2 \pi r dr =  \frac{2 \, _2F_3\left(\frac{1}{2},\frac{3}{2};1,1,\frac{5}{2};-z_1^2\right)}{3 J_1(z_1){}^2} R  \simeq 0.424 \times R \, .
\end{equation}
\noindent where $_{2}\!F_{3}$  is the generalized hypergeometric function.\\ 
\noindent For the disk, also note that the drift term of the taboo process is

\begin{equation}
	\frac{\bm{ \nabla} \varphi_1(r)}{\varphi_1(r)} =- \frac{z_1}{R} \frac{J_1\left( z_1 \frac{r}{R} \right)}{J_0\left( z_1 \frac{r}{R} \right)} \, ,
\end{equation}

\noindent as found in~\cite{ref_Garbaczewski}.

\newpage
%%%%%%%%%%%%%%%%%%%%%%%%%%%%%%%%%%%%%%%%%%%%%%%%%%%%%%%%%%%%%%%%%%%%%%%%%%%%%%%%%%%%%%%%%%%%%%%%%%%%%%%%%%%%%%%%
%%																						%%
%%                BIBLIOGRAPHY                                                                                %%
%%																						%%
%%%%%%%%%%%%%%%%%%%%%%%%%%%%%%%%%%%%%%%%%%%%%%%%%%%%%%%%%%%%%%%%%%%%%%%%%%%%%%%%%%%%%%%%%%%%%%%%%%%%%%%%%%%%%%%%

\end{document}